\documentclass[a4paper,10pt]{article}
\usepackage{fullpage}
\usepackage{amsmath}
\usepackage{epsfig}
\usepackage{amssymb}
\usepackage{amsfonts}
\usepackage{epsfig}
\usepackage{color}
\usepackage{color}

\newcommand{\field}[1]{\mathbb{#1}}
\newcommand{\nn}{\nonumber}

\def\a{\alpha}
\def\b{\beta}
\def\g{\gamma}
\def\d{\delta}

\def\z{\zeta}
\def\q{\theta}
\def\l{\lambda}
\def\s{\sigma}

\def\t{\tau}

\def\c{\chi}
\def\w{\omega}

\def\D{{\cal D}}
\def\G{{\cal G}}
\def\L{{\cal L}}
\def\M{{\cal M}}
\def\O{{\cal O}}
\def\R{{\cal R}}
\def\S{{\cal S}}

\def\Z{{\cal Z}}

\def\Nt{\tilde{N}}
\def\st{\tilde{\sigma}}
\def\cb{\bar{\chi}}
\def\Nb{\bar{N}}
\def\sb{\bar{\sigma}}
\def\wb{\bar{\omega}}
\def\ch{\hat{\chi}}
\def\Nh{\hat{N}}
\def\sh{\hat{\sigma}}
\def\SR{^S\!\bar{R}}
\def\Gt{{\tilde{G}}}
\def\Lt{{\tilde{\Lambda}}}
\def\Tb{\bar{T}}

\def\gt{\tilde{g}}

\def\p{\partial}

\newcommand{\be}{\begin{equation}}
\newcommand{\ee}{\end{equation}}
\newcommand{\bea}{\begin{eqnarray}}
\newcommand{\eea}{\end{eqnarray}}

\newcommand{\cF}{\mathcal F}
\newcommand{\cM}{\mathcal M}
\newcommand{\cO}{\mathcal O}
\newcommand{\cG}{\mathcal G}

\begin{document}
\thispagestyle{empty}
\begin{flushright} \small
\mbox{}
\end{flushright}
\bigskip

\begin{center}
 {\LARGE\bfseries  
 Renormalization group flow of \\[1.5ex]
Ho\v{r}ava-Lifshitz gravity at low energies  
}
\\[12mm]
Adriano Contillo$^1$, Stefan Rechenberger$^2$ and Frank Saueressig$^1$ \\[3mm]
{\small\slshape
$^1$ Institute for Mathematics, Astrophysics and Particle Physics (IMAPP),\\
Radboud University Nijmegen, Heyendaalseweg 135, 6525 AJ Nijmegen, The Netherlands \\[2ex]
$^2$ Institut f\"ur Kernphysik (Theoriezentrum),\\
Technische Universit\"at Darmstadt, D-64289 Darmstadt, Germany\\[2ex]
{\upshape\ttfamily a.contillo@science.ru.nl} \\
{\upshape\ttfamily stefan.rechenberger@physik.tu-darmstadt.de} \\
{\upshape\ttfamily f.saueressig@science.ru.nl} }\\
\end{center}
\vspace{10mm}

\hrule\bigskip

\centerline{\bfseries Abstract} \medskip \noindent
The functional renormalization group equation for projectable Ho\v{r}ava-Lifshitz gravity is used to derive the non-perturbative beta functions for the Newton's constant, cosmological constant and anisotropy parameter. The resulting coupled differential equations are studied in detail and exemplary RG trajectories are constructed numerically. The beta functions possess 
a non-Gaussian fixed point and a one-parameter family of Gaussian fixed points. One of the Gaussian fixed points corresponds to the Einstein-Hilbert action with vanishing cosmological constant and constitutes a saddle point with one IR-attractive direction. For RG trajectories dragged into this fixed point at low energies diffeomorphism invariance is restored. The emergence of general relativity from Ho\v{r}ava-Lifshitz gravity can thus be understood as a crossover-phenomenon where the IR behavior of the theory is controlled by this Gaussian fixed point. In particular RG trajectories with a tiny positive cosmological constant also come with an anisotropy parameter which is compatible with experimental constraints, providing a mechanism for the approximate restoration of diffeomorphism invariance in the IR. The non-Gaussian fixed point is UV-attractive in all three coupling constants. Most likely, this fixed point is the imprint of Asymptotic Safety at the level of Ho\v{r}ava-Lifshitz gravity.

\noindent

\bigskip
\hrule\bigskip
\newpage

\section{Introduction}
Ho\v{r}ava-Lifshitz (HL) gravity \cite{Horava:2008ih,Horava:2009uw,Horava:2009if} (reviewed in, \emph{e.g.}, \cite{Weinfurtner:2010hz,Horava:2011gd,Visser:2011mf}) constitutes an attempt to find a consistent quantum theory of gravity that is perturbatively renormalizable. The key ingredient of the formulation is the breaking of Lorentz invariance at high energies. In this course, the construction assumes a preferred foliation of spacetime into one time direction and three-dimensional spatial slices. At the same time the underlying symmetry group is restricted from the full diffeomorphism group to foliation-preserving diffeomorphisms. This modification naturally leads to anisotropic gravity models which contain a different number of derivatives in time and space directions. The fundamental action underlying HL gravity then contains a kinetic piece which is second order in the time derivatives and potential terms which contain higher order in the spatial derivatives 
\be\label{hlprop}
S = \frac{1}{16 \pi G} \int d\tau d^3x \sqrt{\sigma} N \left( K_{ij} K^{ij} - \lambda K^2 + V[\sigma_{ij}] \right) \, . 
\ee
Here the four-dimensional metric has been decomposed into the lapse function $N$, the shift vector $N_i$ and the metric on the spatial slice $\sigma_{ij}$. The kinetic term is constructed from the extrinsic curvature of the foliation $K_{ij}$ and carries the free parameter $\l$. The potential $V[\sigma]$ is constructed from the intrinsic curvatures (and their derivatives) on the spatial slice. For anisotropic scaling $z$, $V$ includes $2 z$ derivatives. Based on \eqref{hlprop} there are two versions of HL gravity studied in the literature. The projectable case assumes that $N = N(t)$ is independent of the spatial coordinates while in the non-projectable case, $N = N(t,x)$, the lapse function depends on space and time. In the following we will consider the projectable case only, since the non-projectable version seems to suffer from technical inconsistencies \cite{Henneaux:2009zb,Farkas:2010dw}.

In principle, the potential $V[\sigma]$ should include all power-counting relevant and marginal operators that are compatible with foliation-preserving diffeomorphisms \cite{Horava:2009uw}. In 3+1 spacetime dimensions this includes all interaction terms with up to six powers of the spatial derivatives, yielding four relevant and five marginal interaction terms. In \cite{Horava:2009uw} it was also suggested to restrict the number of coupling constants appearing in the model by imposing a detailed balance condition, i.e., demanding that the potential $V[\sigma]$ can be written as the gradient flow of some ``superpotential''. This does, however, not constitute a stringent requirement for the construction to work and may be relaxed \cite{Horava:2009uw,Sotiriou:2009gy}.

 The key property of the construction is that the higher powers of spatial derivatives modify the ultraviolet behavior of the propagator  $G(\omega, {\bf p})$ in the theory. In the deep ultraviolet (UV) this takes the schematic form 
\be
G(\omega, {\bf p}) \propto \left( \omega^2 - \G ({\bf p}^2)^z \right)^{-1} \, , 
\ee
where $z = 3$ is the dynamical critical exponent, $\G$ a coupling constant, and $\omega$ and $\bf p$ denote the momentum in time and spatial directions, respectively. The presence of the higher derivative spatial terms then softens the UV divergences of loop diagrams and it has been argued \cite{Visser:2009fg} that the theory is power-counting renormalizable. 

A serious problem haunting HL gravity is the existence of an additional scalar degree of freedom, not present in classical general relativity. In general this mode turns out to be unstable. Moreover, once the parameters of the theory are adjusted in such a way that its classical low energy limit corresponds to general relativity, the scalar exhibits strong coupling 
\cite{Charmousis:2009tc,Blas:2009yd}. This feature makes it challenging to reconcile HL gravity with phenomenological observations at low energies, where Lorentz symmetry has been tested with very high accuracy \cite{Mattingly:2005re}. Thus, at this stage there is no convincing mechanism from which classical general relativity emerges dynamically in the infrared (IR). Moreover, while the theory is power-counting renormalizable, the existence of a suitable anisotropic UV fixed point that renders the theory perturbatively renormalizable has not been established. Clearly, answering these questions requires a detailed study of the theories RG flow. Despite a wash of papers these central issues have not been settled so far, see however \cite{Orlando:2009en,Briscese:2012rz} for related discussions and preliminary results.\footnote{For further works on the renormalization
of anisotropic quantum field theories also see \cite{Anselmi:2007ri,Anselmi:2008ry,Iengo:2009ix,Giribet:2010th,LopezNacir:2011mt}.}

A practical tool for addressing these questions in a systematic way is the functional renormalization group equation (FRGE) for the effective average action $\Gamma_k$ \cite{avact}. This equation has already been highly successful for understanding RG flows in gauge theories \cite{ym,avactrev,ymrev} and its formulation for metric gravity \cite{mr} has played a central role in developing the gravitational Asymptotic Safety program \cite{Niedermaier:2006wt,Reuter:2007rv,Percacci:2007sz,Litim:2008tt,Reuter:2012id}. 
The effective average action $\Gamma_k$ (together with its FRGE) for projectable HL gravity has been constructed recently in \cite{Manrique:2011jc,Rechenberger:2012dt}. Approximating
$\Gamma_k$ by the ADM-decomposed Einstein-Hilbert action, it was shown that the resulting RG flow matches the one found in metric gravity. In addition the non-Gaussian fixed point, being
the central constituent of the gravitational Asymptotic Safety program, was shown to be robust with respect to a change of spacetime signature \cite{Manrique:2011jc}.

In this work we use this FRGE to compute the beta functions governing the scale dependence of the coupling constants of HL gravity in the low energy approximation. For this purpose, we truncate the potential $V[\sigma]$ appearing in $\Gamma_k$ to the two lowest-dimensional operators
\be\label{potential}
V[\sigma] \approx - R + 2 \Lambda \, . 
\ee
Here $R$ is the scalar curvature on the spatial slice, constructed from $\sigma$, and $\Lambda$ denotes the dimensionful cosmological constant. In the language of RG flows this ansatz constitutes the simplest extension of the Einstein-Hilbert truncation to the case of anisotropic gravity theories. We stress that the computation is not a classical RG flow where one studies, e.g., the limiting behavior of
the classical propagator in the high- and low-energy approximation. Instead we determine the scale dependence of the coupling constants $G$, $\Lambda$ and the anisotropy parameter $\lambda$ induced by the quantum fluctuations of the foliated spacetime. To the best of our knowledge, our computation constitutes the first study of RG flows within HL gravity that includes a coupling constant associated with broken diffeomorphism invariance. As a central result, we establish that the beta functions encoding the scale dependence of $G, \Lambda$, and $\lambda$ possess a Gaussian fixed point whose associated action is given by the Einstein-Hilbert action with zero cosmological constant. This fixed point exhibits an IR-attractive eigendirection (essentially being aligned with Newton's constant) and can thus act as an IR completion of the theory where Lorentz invariance is restored dynamically.

The remainder of this paper is organized as follows. In Section \ref{sec:RGeqn} we briefly review the ADM decomposition together with the construction of the Wetterich equation for projectable HL gravity. This provides the starting point for the derivation of the beta functions capturing the scale dependence of the action \eqref{hlprop} with $V[\sigma]$ given by \eqref{potential} in Section \ref{sec:betas}. The flow generated by these beta functions is studied in Sections \ref{sec:RGflow} and \ref{sec:flowII}, with a particular emphasis on the low energy regime of the theory. In Section \ref{sec:con} we discuss our results and draw our conclusions.

\section{The Wetterich equation for foliated spacetimes}
\label{sec:RGeqn}
We start our analysis by briefly summarizing the classical ADM decomposition of the metric field \cite{Arnowitt:1959ah}
and the construction of the FRGE for projectable HL gravity \cite{Rechenberger:2012dt}, which provide the foundation
for our analysis. A more
detailed discussion can be found in \cite{Rechenberger:2012dt}.

\subsection{Foliated spacetimes} 
\label{subsec:fol}
The ADM construction starts from a four-dimensional manifold $\M$ with metric $\g_{\a\b}$. In principle, the spacetime signature can either be Euclidean $(+,+,+,+)$, or Lorentzian $(-,+,+,+)$. In the sequel, we restrict ourselves to Euclidean signature. Given the foliation structure implied by the ADM decomposition, the Lorentzian case can be obtained by a Wick rotation.

Introducing the lapse function $\tilde N$, the shift vector $\tilde N_i$ and the induced metric on the spatial slices $\st_{ij}$ the infinitesimal square line element becomes
\begin{equation}
 \mathrm ds^2=\g_{\a\b}\mathrm dx^\a \mathrm dx^\b=\Nt^2\mathrm d\t^2+\st_{ij}\left(\mathrm dy^i+\Nt^i\mathrm d\t\right)\left(\mathrm dy^j+\Nt^j\mathrm d\t\right)
\end{equation}
where the component fields depend on the spacetime coordinates $\left\{\Nt(\t,y),\Nt_i(\t,y),\st_{ij}(\t,y)\right\}$. From this we can derive the expression of the spacetime metric $\g_{\a\b}$ and its inverse $\g^{\a\b}$ as a function of the ADM fields
\begin{equation}\label{ADMdecomp}
 \g_{\a\b}=\left(\begin{array}{cc}
 \Nt^2+\Nt_i\Nt^i & \Nt_j \\
 \Nt_i & \st_{ij}\end{array}\right) \, , \qquad
 \g^{\a\b}=\left(\begin{array}{cc}
\frac{1}{\Nt^2} & -\frac{\Nt^j}{\Nt^2} \\
-\frac{\Nt^i}{\Nt^2} & \st^{ij}+\frac{\Nt^i\Nt^j}{\Nt^2}\end{array}\right)\;.
\end{equation}
Here the scalar products are with respect to the spatial metric $\st_{ij}$. For completeness, we note that the decomposition of the determinant, appearing in the spacetime volume, is given as $\sqrt{\g}=\Nt\sqrt{\st}$.

Under a general coordinate transformation Diff($\M$) the infinitesimal variation of the metric field can be written as
\begin{equation}\label{liederiv}
 \d\g_{\a\b}=\L_v\g_{\a\b} \, ,
\end{equation}
$\L_v$ being the Lie derivative with respect to the four-dimensional vector $v^\a$. Rewriting $v^\a$ in terms of its temporal and spatial parts, $f$ and $\z_i$, as $v^\a=(f(\t,x),\z^i(\t,x))$, the transformation of the component fields under Diff($\M$) reads
\begin{align}\label{gaugevars}
\d\Nt&=\p_\t(f\Nt)+\z^k\p_k\Nt-\Nt\Nt^i\p_if\,,\nn\\
\d\Nt_i&=\p_\t(\Nt_if)+\L_\z(\Nt_i) +\st_{ki}\p_\t\z^k+\Nt_k\Nt^k\p_if+\Nt^2\p_if\,,\\
\d\st_{ij}&=f\p_\t\st_{ij}+\L_\z(\st_{ij})+\Nt_j\p_if+\Nt_i\p_jf\;.\nn
\end{align}
For completeness, we note
\begin{equation}\label{Nui}
\d\Nt^i=\p_\t(\Nt^if)+\L_\z(\Nt^i) +\p_\t\z^i-\Nt^i\Nt^j\p_jf+\Nt^2\st^{ij}\p_jf\;.
\end{equation}
Here $\L_\z$ denotes the Lie derivative on the spatial slices which contains only spatial derivatives.
One observes that, while Diff($\M$) acts linearly on the metric $\g_{\a\b}$, the nonlinearity of the ADM decomposition (\ref{ADMdecomp}) leads to a nonlinear transformation law for the component fields. The freedom of choosing a coordinate system can be exploited to (locally) adopt the proper-time gauge \cite{Dasgupta:2001ue}
\begin{equation}
 \Nt(\t,x)=1\quad,\qquad\Nt_i(\t,x)=0\;.
\end{equation}

Ho\v{r}ava-Lifshitz gravity \cite{Horava:2008ih,Horava:2009uw,Horava:2009if} encodes the gravitational degrees of freedom in terms of the ADM fields. Considering the projectable version of the theory, the key difference is that here only the metric $\st_{ij}$ on the spatial slices $\Sigma$ and the shift vector $\Nt_i$ are spacetime-dependent fields, while the lapse function $\Nt(\t)$ depends on time only and is constant along $\Sigma$. The component fields of this projectable version thus read
\begin{equation}\label{projfieldcontent}
 \tilde{\c}^{\textrm{pHL}}=\left\{\Nt(\t),\Nt_i(\t,y),\st_{ij}(\t,y)\right\}\;.
\end{equation}
Moreover, the symmetry group is restricted to foliation-preserving diffeomorphisms Diff($\M$,$\Sigma$). In this case the vector appearing in the Lie derivative (\ref{liederiv}) is restricted to the form
\begin{equation}
 v^\a=(f(\t),\z^i(\t,y))\;.
\end{equation}
The gauge variations (\ref{gaugevars}) therefore acquire the simplified form
\begin{equation}\label{projgaugevars}
\begin{split}
\d\Nt&=\p_\t(f\Nt)\,,\\
\d\Nt_i&=\p_\t(f\Nt_i)+\L_\z(\Nt_i) + \st_{ik}\p_\t\z^k\,,\\
\d\st_{ij}&=f\p_\t\st_{ij}+\L_\z(\st_{ij}) 
\end{split}
\end{equation}
and
\begin{equation}
\d\Nt^i=\p_\t(f\Nt^i)+\L_\z(\Nt^i)+\p_\t\z^i\, . 
\end{equation}
In a sense, projectable HL gravity may thus be understood as a partially gauge-fixed version of the general ADM decomposition. 
The reduced symmetry requirements permit interaction terms which are forbidden within a fully diffeomorphism-invariant theory.
Thus projectable HL gravity constitutes a genuine generalization of metric formulation of gravity. 

\subsection{The functional RG equation}
Given the field content and the symmetry requirements
of the theory, we can proceed with the construction of the 
Wetterich equation for projectable HL gravity. The starting point
is the path integral including a generic 
gravitational action $\S^{\textrm{grav}}[\Nt,\Nt_i,\st_{ij}]$
constructed from the fields \eqref{projfieldcontent} and invariant
under \eqref{projgaugevars}. 

In order to make the path integral well defined, the symmetries of the
theory have to be gauge-fixed. Following the construction
of the FRGE for metric gravity \cite{mr}, this is done through
the background field method reviewed in \cite{Niedermaier:2006wt}. In this case
 the fluctuation fields $\tilde{\c}$
are split into a fixed but arbitrary background component $\cb=\{\Nb,\Nb_i,\sb_{ij}\}$ and quantum fluctuations $\ch=\{\Nh,\Nh_i,\sh_{ij}\}$ such that
\begin{equation}
 \tilde{\c}\equiv\cb+\ch\;.
\end{equation}
The fluctuations are arbitrary and not restricted to be small. Quantum gauge transformations
then attribute the transformations \eqref{projgaugevars} to the fluctuation fields, keeping the background
fields untouched
\be
\delta_Q \cb \equiv 0 \, , \qquad \delta_Q \ch \equiv \delta \tilde{\c} \, . 
\ee
In addition, the background field method allows to introduce so-called background gauge transformations acting on both components, in such a way that the transformation of the background fields becomes identical to the one of the quantum fields once the fluctuations are set to zero. In general these transformations are not determined uniquely and we chose
\begin{equation}
\begin{split}
 \d_B\Nb&=\p_\t(f\Nb) \, , \\
 \d_B\Nb_i&=\p_\t(f\Nb_i)+\st_{ij}\p_\t\z^j+\L_\z(\Nb_i) \, , \\
 \d_B\sb_{ij}&=f\p_\t\sb_{ij}+\L_\z(\sb_{ij}) \, , 
\end{split}
\end{equation}
for the background and
\begin{equation}
\begin{split}
 \d_B\Nh&=\p_\t(f\Nh) \, ,  \\
 \d_B\Nh_i&=\p_\t(f\Nh_i)+\L_\z(\Nh_i) \, ,  \\
 \d_B\sh_{ij}&=f\p_\t\sh_{ij}+\L_\z(\sh_{ij}) \, , 
\end{split}
\end{equation}
for the fluctuations.

We can now construct a gauge-fixing that implements the proper-time gauge in the background field formalism. The corresponding gauge-fixing
action has to break the quantum gauge transformations while respecting the background gauge transformations. These requirements
can be met by setting
\begin{equation}\label{gfaction}
 \S^{\textrm{gf}}=\frac{1}{2}\int \mathrm d\t \mathrm d^3\!x\,\Nb\sqrt{\sb}\,\left(\frac{\Nh^2}{\a_L\Nb^2}+\frac{\sb_{ij}\Nh^i\Nh^j}{\a_S\Nb^2}\right) \, . 
\end{equation}
In the Landau limit $\a_L,\a_S\to0$, $\S^{\textrm{gf}}$ becomes a delta distribution which eliminates the fluctuations of the lapse function and the shift vector.
This gauge choice actually differs from the one advocated by Ho\v{r}ava \cite{Horava:2009uw}, which implements a harmonic constraint on the metric fluctuations on
the spatial slice instead of fixing $\hat{N}$. For practical computations of the RG flow the present choice is more convenient, however, since it avoids the appearance of off-diagonal fluctuation terms in the FRGE.

The ghost action arising from the choice \eqref{gfaction} can then be found in the standard way
\begin{equation}\label{ghostaction}
 \S^{\textrm{gh}}=\int \mathrm d\t \mathrm d^3\!x\,\Nb\sqrt{\sb}\,\left[\wb\,\p_\t\frac{\Nt}{\Nb}\,\w+\wb_i\left(\d^i_j\p_\t-\d^i_j\Nt^k\p_k+\p_j\Nt^i\right)\w^j\right] \, . 
\end{equation}
Here the vector ghosts $\w^i(\t,x)$ and $\wb_i(\t,x)$ are functions of the spacetime, while the scalar ghosts $\w(\t)$ and $\wb(\t)$ depend on time only. The background lapse function $\Nb(\t)$ has been distributed in such a way that the ghost action is invariant under background gauge transformations with the ghosts transforming respectively as scalars and vectors
\begin{equation}
\begin{split}
 \d_B\w=f\p_\t\w \, , \quad & \quad\d_B\wb=f\p_\t\wb \,  , \\
 \d_B\w^i=f\p_\t\w^i+\L_\z(\w^i)\quad \, ,  & \quad\d_B\wb_i=f\p_\t\wb_i+\L_\z(\wb_i)\;.
\end{split}
\end{equation}

Given these prerequisites, the construction of the effective average action $\Gamma_k$ for projectable HL gravity together with its
FRGE follows the usual procedure. The scale-dependent generating functional of connected Green functions is given by
\begin{equation}
 \exp\left\{W_k[J;\cb]\right\}=\int\D\hat{\mu}\, \exp\left[-\S^{\textrm{grav}}-\S^{\textrm{gf}}-\S^{\textrm{gh}}-\Delta_k\S-\S^{\textrm{source}}\right]
\end{equation}
where $\S^{\textrm{grav}}[\Nt,\Nt_i,\st_{ij}]$ is a generic gravitational action, $J$ indicates the source terms for the fluctuation fields and the measure $\D\hat{\mu}$ consists of the integration over the gravitational and ghost fluctuations. The crucial ingredient is the scale-dependent IR cutoff term $\Delta_k\S$, which separates fluctuations
into IR ($p^2 < k^2$) and UV ($p^2 > k^2$) modes. This cutoff acts as a mass term for the IR fluctuations, suppressing their contribution to the pathintegral.
Formally, it can be written as

\begin{equation}\label{cutoffterm}
\begin{split}
 \Delta_k\S&=\frac{1}{2}\int \mathrm d\t \mathrm d^3\!x\,\Nb\sqrt{\sb}\,\ch\,\R^{\textrm{grav}}_k[\cb]\,\ch 
 +\int \mathrm d\t \mathrm d^3\!x\,\Nb\sqrt{\sb}\,(\w,\w_i)\,\R^{\textrm{gh}}_k[\cb]\,(\wb,\wb_j)^{\textrm{T}}\; .
\end{split}
\end{equation}
The cutoff operators $\R_k$ are matrix valued in field space and follow the general structure
\begin{equation}\label{cutoffstructure}
 \R_k = \Z_k R_k = \Z_k\,k^2\,R^{(0)}(p^2/k^2)
\end{equation}
where $\Z_k$ is a background-field-dependent matrix that ensures the invariance of (\ref{cutoffterm}) with respect to the background gauge transformations and $p$ indicates the mode momentum.
The profile function $R^{(0)}$ interpolates between $\lim_{z \rightarrow 0} R^{(0)}(z) = 1$ 
and $\lim_{z \rightarrow \infty} R^{(0)}(z) = 0$. In practical computations, it is often convenient to
work with the optimized cutoff \cite{opt}, setting $R^{(0)} = (1-z)\theta(1-z)$ with $\theta$ the Heaviside step function.

The effective average action $\Gamma_k$ is given by the scale-dependent Legendre transform of $W_k$
\begin{equation}
 \Gamma_k[\c;\cb]=\S^{\textrm{source}}-W_k[J;\cb]-\Delta_k\S \, . 
\end{equation}
Taking the derivative with respect to the RG time $t=\ln(k/k_0)$, it has been shown in \cite{Rechenberger:2012dt} that the scale dependence of $\Gamma_k$
is encoded in a Wetterich-type FRGE
\begin{equation}\label{wettericheqn}
 \p_t\Gamma_k=\frac{1}{2}\,\textrm{STr}\left[\left(\Gamma^{(2)}_k+\R_k\right)^{-1}\p_t\R_k\right] \, .
\end{equation}
Here $\Gamma^{(2)}_k$ denotes the second derivative of $\Gamma_k$ with respect to the fluctuation fields $\hat{\c}$ and the supertrace includes an integral over loop momenta, a trace in field space and a minus sign for ghost contributions. This expression constitutes the desired functional renormalization group equation for projectable Ho\v{r}ava-Lifshitz gravity.

\section{The beta functions of Ho\v{r}ava-Lifshitz gravity}
\label{sec:betas}

The Wetterich equation (\ref{wettericheqn}) is a very powerful tool for the study of RG flows. A common technique to find approximate solutions of the equation which do not rely on an expansion in a small parameter consists in truncating the effective average action by restricting $\Gamma_k$ to a finite set of running coupling constants. In the following the ansatz for $\Gamma_k$ will be of the general form
\begin{equation}\label{actionansatz}
 \Gamma_k[\c;\cb]\simeq\Gamma^{\textrm{grav}}_k[\c;\cb]+\S^{\textrm{gf}}[\c;\cb]+\S^{\textrm{gh}}[\c;\cb]
\end{equation}
where we approximated the gauge-fixing and ghost parts by their classical expressions (\ref{gfaction}) and (\ref{ghostaction}).

The procedure followed here departs from the one of \cite{Rechenberger:2012dt}, in order to achieve the different aims of the two analyses. The main aim of \cite{Rechenberger:2012dt} was to compare the results of the Einstein-Hilbert flow in the metric and foliated approaches. We are here going to use the latter, but we are now more interested in the possibility of a Lorentz-symmetry breaking RG flow. This is why, when rewriting the Einstein-Hilbert action
\begin{equation}
 \Gamma^{\textrm{grav}}_k=\frac{1}{16\pi G_k}\int \mathrm d^4\!x\,\sqrt{g}\,\left[-{}^{(4)}R+2\Lambda_k\right]
\end{equation}
in terms of the lapse function, the shift vector and the spatial metric, we introduce an additional coupling constant $\l_k$ that will keep track of the spacetime anisotropy. The existence of this new coupling constant is a direct consequence of the reduced degree of symmetry that we are imposing to our system, but the actual position inside the action is chosen to match the one given in \cite{Horava:2008ih,Horava:2009uw}. The new expression for the gravitational action will therefore be
\begin{equation}\label{ADMaction}
 \Gamma^{\textrm{grav}}_k=\frac{1}{16\pi G_k}\int \mathrm d\t \mathrm d^3\!x\,N\sqrt{\s}\,\left[K_{ij} \, \cG^{ikjl} \, K_{kl} - {}^{(3)\!\!}R+2\Lambda_k\right]\;.
\end{equation}
Here
\begin{equation}
 K_{ij}=\frac{1}{2N}\left(\p_\t\s_{ij}-D_iN_j-D_jN_i\right)
\end{equation}
is the extrinsic curvature and the kinetic term is constructed from the running Wheeler-de Witt metric
\be
\cG^{ikjl} \equiv \sigma^{ik} \sigma^{jl} - \lambda_k \, \sigma^{ij} \sigma^{kl} \, , 
\ee
containing the scale-dependent anisotropic coupling $\l_k$. The fields $\c = \left\{N,N_i,\s_{ij}\right\}$ are the classical counterparts of the full quantum fields $\tilde{\c}$. Furthermore, $D_i$ and $^{(3)}R$ denote the covariant derivative and
the intrinsic curvature of the three-dimensional spatial slices constructed from $\s_{ij}$. 
The coupling $\lambda_k$ serves as a good indicator for breaking the full diffeomorphism invariance of the theory, since the relative weight of the kinetic terms can not be absorbed by a rescaling of the fields. The action encompasses
the two interesting cases $\lambda = 1$ and $\lambda = 1/3$.  In the first case \eqref{ADMaction} coincides with
the Einstein-Hilbert action. The second value is special since for this case the kinetic term for the 
fluctuations of the conformal mode decouples so that \eqref{ADMaction} becomes scale invariant at the classical level.

\subsection{Constructing the functional traces}
Inserting the ansatz (\ref{actionansatz}) into the Wetterich equation (\ref{wettericheqn}), we can read off the scale dependence of the couplings from the coefficients of the spacetime volume as well as the intrinsic and extrinsic curvature of the spatial slices. Thus it is sufficient to project the traces appearing on the right hand side onto these curvature invariants. We stress that when computing the beta functions of the theory the geometric quantities merely act as bookkeeping devices. No physical meaning should be attached to them.

The first step in evaluating the functional traces constitutes in computing the Hessian $\Gamma^{(2)}_k$ with respect to the fluctuation fields. This calculation can be simplified in a number of ways. First we adopt Landau gauge by taking the limit $\a_L,\a_S\to0$. In this limit the gauge-fixing term (\ref{gfaction}) is converted to a delta functions which freeze the fluctuation fields $\Nh=0$, $\Nh_i=0$. Thus $\Nh$ and $\Nh_i$ decouple and do not contribute to the traces on the right hand side of the flow equation. Therefore we are left with the background fields $\bar N, \bar N_i$ and the classical spatial metric
\begin{equation}
\s_{ij}=\sb_{ij}+h_{ij}
\end{equation}
with the background $\sb_{ij}$ and the fluctuations $h_{ij}$.

The second simplification comes from selecting a class of special backgrounds. This choice is almost arbitrary since the background is only a technical tool. The guiding principle is that it should be as simple as possible to enable us to distinguish the different interaction monomials. A natural choice is
\begin{equation}\label{backgroundmanifold}
 \M=S^1\times S^3_{\textrm{warped}}
\end{equation}
\emph{i.e.} the direct product of a time circle with periodicity $T$ and a ``warped'' three-sphere with time-dependent curvature radius. In terms of the background multiplet this choice implies
\begin{equation}\label{tdepbackground}
 \Nb(\t)=1\;,\quad\Nb_i(\t,x)=0\;,\quad\sb_{ij}(\t,x)=\c(\t)\bar{s}_{ij}
\end{equation}
where $\bar{s}_{ij}$ indicates the spatial background metric of the time-independent three-sphere, whose curvature invariants satisfy
\begin{equation}
 \p_\t\bar{s}_{ij}=0\;,\quad \SR_{ij}=\frac{\SR}{3}\bar{s}_{ij}\;,\quad \SR_{ijkl}=\frac{\SR}{6}(\bar{s}_{ik}\bar{s}_{jl}-\bar{s}_{il}\bar{s}_{jk})
\end{equation}
due to its maximal symmetry. The index $S$ marks the time-independent sphere and we dropped the prefix $(3)$ of the intrinsic curvature in order to lighten our notation. 
The $\tau$ dependence of the background metric has the consequence that time derivatives no longer commute with raising and lowering spatial indices with $\sb$. In order to
lift this potential ambiguity, we adopt the notation that the background metric used to raise the indices of the fluctuation fields \emph{are to the left} of all time-derivatives, i.e.,
$\int \mathrm d\tau \mathrm d^3x \sqrt{\sb} \, h^{ij} \p_\tau^2 h_{ij} \equiv \int \mathrm d\tau \mathrm d^3x \sqrt{\sb} \, h_{kl} \sb^{ki} \sb^{lj} \p_\tau^2 h_{ij}$, etc.

Substituting this particular choice of background and setting $\c=\bar{\c}$ afterwards, the ansatz (\ref{ADMaction}) simplifies to
\begin{equation}\label{prolhs}
 \left.\Gamma_k^{\textrm{grav}}\right|_{\bar{\c}}=\frac{1}{16\pi G_k}\int \mathrm d\t \mathrm d^3\!x\,\sqrt{\bar{s}}\,\c^{3/2}\,\left[
\frac{3\g^2}{4}(1-3\l_k)-\frac{\SR}{\c}+2\Lambda_k\right]
\end{equation}
where $\g=\p_\t\ln\c$. Note that both parts of the kinetic term are proportional to $\g^2$. Thus the background allows us to distinguish three couplings. The cosmological constant $\Lambda_k$ corresponds to the volume term, the Newton's constant $G_k$ is related to the scalar curvature term and the last coupling, $\l_k$, is associated with the parameter $\g^2$.

We then proceed to compute the Hessian of $\Gamma^{\textrm{grav}}_k$ around the background (\ref{backgroundmanifold}). Inspecting (\ref{ADMaction}), there are four distinguished interaction monomials
\begin{align}
 &I_1=\int \mathrm d\t \mathrm d^3\!x\,\sqrt{\s} \, , && I_3=\int \mathrm d\t \mathrm d^3\!x\,\sqrt{\s}\,K_{ij}\s^{ik}\s^{jl}K_{kl} \, , \nonumber \\
 &I_2=\int \mathrm d\t \mathrm d^3\!x\,\sqrt{\s}\,R \, , && I_4=\int \mathrm d\t \mathrm d^3\!x\,\sqrt{\s}\,K^2\;. 
\end{align}
We evaluate the second variation of these terms and, for convenience, decompose the fluctuation metric $h_{ij}$ into a traceless and trace part according to
\begin{equation}
 h_{ij}=h^{\mathrm T}_{ij}+\frac{1}{3}\sb_{ij}h\;,\quad\sb^{ij}h^{\mathrm T}_{ij}=0\;.
\end{equation}
The result of these variations reads
\begin{align}\label{2ndvariations1}
\delta^2 I_1 &=\int \mathrm d\t \mathrm d^3\!x\,\sqrt{\sb}\,\Big[ - \tfrac{1}{2} \, h^{\mathrm Tij} \, h^\mathrm T_{ij} + \tfrac{1}{12} \, h^2 \Big] \nonumber \\
\delta^2 I_2 &=\int \mathrm d\t \mathrm d^3\!x\,\sqrt{\sb}\,\Big[ \, h \left( \tfrac{1}{9} \Delta - \tfrac{1}{36} \bar R \right) h - \tfrac{1}{2} h^\mathrm T_{ij} \left( \Delta + \tfrac{2}{3} \bar R \right) h^{\mathrm{T}ij} + \tfrac{1}{3} h \bar{D}_i \bar{D}_j h^{\mathrm Tij} - h^{\mathrm Tik} \bar{D}_k \bar{D}_l h^{\mathrm Tlj}\bar\sigma_{ij} \Big] \nonumber \\
\delta^2 I_3 &=\int \mathrm d\t \mathrm d^3\!x\,\sqrt{\sb}\,\Big[ -\tfrac{1}{2} \, h^{\mathrm Tij} \, \partial_\tau^2 \, h^\mathrm T_{ij}  - \tfrac{1}{6} \, h  \, \partial_\tau^2 \, h - \g\left( \tfrac{1}{12} \, h \, \partial_\tau h \, + \, \tfrac{7}{4} \, h^{\mathrm Tij} \, \partial_\tau \, h^\mathrm T_{ij} \right) - \g^2 \left( \tfrac{13}{48} \, h^2 \, - \, \tfrac{1}{8} \, h^{\mathrm Tij} \, h^\mathrm T_{ij} \right) \Big] \nonumber \\
\delta^2 I_4 &=\int \mathrm d\t \mathrm d^3\!x\,\sqrt{\sb}\,\Big[ - \tfrac{1}{2} \, h  \, \partial_\tau^2 \, h \, - \gamma \left( \tfrac{1}{4} \, h \, \partial_\tau h \, + \, 3 \, h^{\mathrm Tij} \, \partial_\tau \, h^\mathrm T_{ij} \right) - \g^2 \left( \tfrac{13}{16} \, h^2 \, + \, \tfrac{9}{8} \, h^{\mathrm Tij} \, h^\mathrm T_{ij} \right) \Big]
\end{align}
where $\bar{D}_i$ denotes the covariant derivative with respect to $\sb_{ij}$, while $\Delta=-\sb^{ij}\bar{D}_i\bar{D}_j$ is the corresponding Laplace operator.

Because of the presence of a non-diagonal term in $\d^2I_2$, leading to a non-minimal operator structure in $\Gamma^{(2)}_k$, it is necessary to further decompose the traceless part of the metric fluctuations $h^{\mathrm T}_{ij}$ into its irreducible representations on the sphere via the transverse traceless (TT) decomposition \cite{York}
\begin{equation}\label{TTdecomp}
h^\mathrm T_{ij} = h^\mathrm{TT}_{ij} + \bar\nabla_i\xi_j + \bar\nabla_j\xi_i + \bar\nabla_i\bar\nabla_j\varsigma + \tfrac{1}{3}\bar\sigma_{ij}\Delta\varsigma \, .
\end{equation}
The component fields are subject to the differential constraints
\begin{equation}
 \sb^{ij}h^{\mathrm{TT}}_{ij}=0\;,\quad\bar{D}^ih^{\mathrm{TT}}_{ij}=0\;,\quad\bar{D}^i\xi_i=0\;.
\end{equation}
Applying this split we find a block-diagonal form of the second variations. Explicitly they read
\begin{align}\label{eq:secondVariations2}
\delta^2 I_1 &=\int \mathrm d\t \mathrm d^3\!x\,\sqrt{\sb}\,\Big\{ - \tfrac{1}{2} h^{\mathrm{TT}ij} h^\mathrm{TT}_{ij} - \xi^i \left[ \Delta - \tfrac{1}{3} \bar R \right] \xi_i - \tfrac{1}{3} \, \varsigma \left[ \Delta^2 - \tfrac{1}{2} \bar R \Delta \right] \varsigma  + \tfrac{1}{12} \, h^2 \Big\} \nonumber \\
\delta^2 I_2 &= \int \mathrm d\t \mathrm d^3\!x\,\sqrt{\sb}\,\Big\{ \tfrac{1}{9} h \left[ \Delta - \tfrac{1}{4} \bar R \right] h - \tfrac{1}{2} h^\mathrm{TT}_{ij} \left[ \Delta + \tfrac{2}{3} \bar R \right] h^{ij\mathrm{TT}}  - \tfrac{1}{3}\bar R \, \xi_i \left[ \Delta - \tfrac{1}{3}\bar R \right] \xi^i \nonumber \\
& \qquad\qquad\qquad\quad + \tfrac{1}{9} \varsigma \Delta\left[ \Delta - \tfrac{1}{2}\bar R \right]\left[ \Delta - \bar R \right]\varsigma + \tfrac{2}{9} \, h \left[ \Delta^2 - \tfrac{1}{2} \bar R \Delta \right] \varsigma \Big\} \nonumber \\
\delta^2 I_3 &=\int \mathrm d\t \mathrm d^3\!x\,\sqrt{\sb}\,\Big\{ -\tfrac{1}{2} h^{\mathrm{TT}ij} \partial_\tau^2 h^\mathrm{TT}_{ij} - \xi^i \left[ \Delta - \tfrac{1}{3} \bar R \right] \partial_\tau^2 \xi_i - \tfrac{1}{3} \, \varsigma \left[ \Delta^2 - \tfrac{1}{2} \bar R \Delta \right] \partial_\tau^2 \varsigma - \tfrac{1}{6} \, h  \, \partial_\tau^2 \, h \nonumber \\
& \qquad\qquad\qquad\quad - \g\left( +\tfrac{1}{12} \, h \, \partial_\tau h \, + \, \tfrac{7}{4} h^{\mathrm{TT}ij} \partial_\tau h^\mathrm{TT}_{ij} + \tfrac{7}{2} \xi^i \left[ \Delta - \tfrac{1}{3} \bar R \right] \partial_\tau \xi_i + \tfrac{7}{6} \, \varsigma \left[ \Delta^2 - \tfrac{1}{2} \bar R \Delta \right] \partial_\tau \varsigma \right) \nonumber \\
& \qquad\qquad\qquad\quad - \g^2 \left( \tfrac{13}{48} \, h^2 \, - \, \tfrac{1}{8} h^{\mathrm{TT}ij} h^\mathrm{TT}_{ij} - \tfrac{1}{4} \xi^i \left[ \Delta - \tfrac{1}{3} \bar R \right] \xi_i - \tfrac{1}{12} \, \varsigma \left[ \Delta^2 - \tfrac{1}{2} \bar R \Delta \right] \varsigma \right) \Big\} \nonumber \\
\delta^2 I_4 &=\int \mathrm d\t \mathrm d^3\!x\,\sqrt{\sb}\,\Big\{ - \tfrac{1}{2} \, h  \, \partial_\tau^2 \, h \nonumber \\
& \qquad\qquad\qquad\quad - \gamma \left( \tfrac{1}{4} \, h \, \partial_\tau h \, + \, 3 h^{\mathrm{TT}ij} \partial_\tau h^\mathrm{TT}_{ij} +6 \xi^i \left[ \Delta - \tfrac{1}{3} \bar R \right] \partial_\tau \xi_i + 2 \, \varsigma \left[ \Delta^2 - \tfrac{1}{2} \bar R \Delta \right] \partial_\tau \varsigma \right) \nonumber \\
& \qquad\qquad\qquad\quad - \g^2 \left( \tfrac{13}{16} \, h^2 \, + \, \tfrac{9}{8} h^{\mathrm{TT}ij} h^\mathrm{TT}_{ij} +\tfrac{9}{4} \xi^i \left[ \Delta - \tfrac{1}{3} \bar R \right] \xi_i + \tfrac{3}{4} \, \varsigma \left[ \Delta^2 - \tfrac{1}{2} \bar R \Delta \right] \varsigma \right) \Big\} \, .
\end{align}
Next, we chose a normalization of the transverse vector and the scalar $\varsigma$ according to
\begin{equation} \label{eq:replacement}
\varsigma \mapsto \sqrt{\Delta} \sqrt{\Delta - \tfrac{1}{2} \bar R} \, \varsigma \, , \qquad \xi^j \mapsto \sqrt{\Delta - \tfrac{1}{3} \bar R} \, \xi^j \, .
\end{equation}
This normalization has the virtue that the Jacobians arising from performing the TT-decomposition are actually canceled by the Jacobi determinants arising 
from this choice of normalization \cite{Lauscher:2001ya}. Thus this construction does not require the introduction of auxiliary fields exponentiating these determinants.
Applying the redefinition \eqref{eq:replacement} the second variations (\ref{eq:secondVariations2}) become
\begin{align}\label{2ndvariations2}
\delta^2 I_1 &=\int \mathrm d\t \mathrm d^3\!x\,\sqrt{\sb}\,\Big\{ \tfrac{1}{12} \, h^2 - \tfrac{1}{2} h^{\mathrm{TT}ij} h^\mathrm{TT}_{ij} - \xi^i \xi_i - \tfrac{1}{3} \, \varsigma^2 \Big\} \nonumber \\
\delta^2 I_2 &= \int \mathrm d\t \mathrm d^3\!x\,\sqrt{\sb}\,\Big\{ \tfrac{1}{9} h \left[ \Delta - \tfrac{1}{4} \bar R \right] h - \tfrac{1}{2} h^\mathrm{TT}_{ij} \left[ \Delta + \tfrac{2}{3} \bar R \right] h^{ij\mathrm{TT}}  - \tfrac{1}{3}\bar R \, \xi_i \xi^i \nonumber \\
& \qquad\qquad\qquad\quad + \tfrac{1}{9} \varsigma \left[ \Delta - \bar R \right]\varsigma + \tfrac{2}{9} \, h \left[ \Delta^2 - \tfrac{1}{2} \bar R \Delta \right]^{1/2} \varsigma \Big\} \nonumber \\
\delta^2 I_3 &=\int \mathrm d\t \mathrm d^3\!x\,\sqrt{\sb}\,\Big\{ - \tfrac{1}{6} \, h  \, \partial_\tau^2 \, h -\tfrac{1}{2} h^{\mathrm{TT}ij} \partial_\tau^2 h^\mathrm{TT}_{ij} - \xi^i \partial_\tau^2 \xi_i - \tfrac{1}{3} \, \varsigma \partial_\tau^2 \varsigma \nonumber \\
& \qquad\qquad\qquad\quad - \g\left( \tfrac{1}{12} \, h \, \partial_\tau h \, + \, \tfrac{7}{4} h^{\mathrm{TT}ij} \partial_\tau h^\mathrm{TT}_{ij} + \tfrac{7}{2} \xi^i \partial_\tau \xi_i + \tfrac{7}{6} \, \varsigma \partial_\tau \varsigma \right) \nonumber \\
& \qquad\qquad\qquad\quad - \g^2 \left( \tfrac{13}{48} \, h^2 \, - \, \tfrac{1}{8} h^{\mathrm{TT}ij} h^\mathrm{TT}_{ij} + \xi^i \xi_i + \tfrac{11}{12} \, \varsigma^2 \right) \Big\} \nonumber \\
\delta^2 I_4 &=\int \mathrm d\t \mathrm d^3\!x\,\sqrt{\sb}\,\Big\{ - \tfrac{1}{2} \, h  \, \partial_\tau^2 \, h \, - \gamma \left( \tfrac{1}{4} \, h \, \partial_\tau h \, + \, 3 h^{\mathrm{TT}ij} \partial_\tau h^\mathrm{TT}_{ij} +6 \xi^i \partial_\tau \xi_i + 2 \, \varsigma \partial_\tau \varsigma \right) \nonumber \\
& \qquad\qquad\qquad\quad - \g^2 \left( \tfrac{13}{16} \, h^2 \, + \, \tfrac{9}{8} h^{\mathrm{TT}ij} h^\mathrm{TT}_{ij} +\tfrac{21}{4} \xi^i \xi_i + \tfrac{11}{4} \, \varsigma^2 \right) \Big\} \, .
\end{align}
Here the implicit time dependence arising from the normalization \eqref{eq:replacement} has explicitly been taken into account.

Based on (\ref{2ndvariations2}) it is now straightforward to write the part of $\Gamma^{\textrm{grav}}_k$ quadratic in the fluctuation fields
\begin{equation}
 \Gamma^{\textrm{grav}}_k[\s_{ij}]=\Gamma^{\textrm{grav}}_k[\sb_{ij}]+\frac{1}{2}\d^2\Gamma^{\textrm{grav}}_k[h_{ij},\sb_{ij}]\;.
\end{equation}
The terms linear in the fluctuations and higher terms are not of interest, since they do not enter into the present computations. It is convenient to list the contributions appearing in the second variation according to the components of the fluctuation fields contained
\begin{align} \label{2ndvariations3}
\delta^2 \Gamma_{\mathrm{TT}}^{\mathrm{grav}} = & \frac{1}{16 \pi G_k} \int \mathrm d\t \mathrm d^3\!x\,\sqrt{\sb}\,\tfrac{1}{2} h^{{\mathrm{TT}}ij} \, \Big[ \Delta + \tfrac{2}{3} \bar R - 2 \Lambda_k - \partial_\tau^2 + \gamma\tfrac{12\lambda_k - 7}{2}\partial_\tau + \gamma^2\tfrac{9\lambda_k + 1}{4} \Big] \, h_{ij}^{\mathrm{TT}} \nonumber \\ \nonumber\\
\delta^2 \Gamma_{\xi\xi}^{\mathrm{grav}} = & \frac{1}{16 \pi G_k} \int \mathrm d\t \mathrm d^3\!x\,\sqrt{\sb}\,\xi^i \Big[ \tfrac{1}{3} \bar R - 2 \Lambda_k - \partial_\tau^2 + \gamma\tfrac{12\lambda_k - 7}{2}\partial_\tau + \gamma^2\tfrac{21\lambda_k - 4}{4} \Big] \xi_i \nonumber \\ \nonumber\\
\delta^2 \Gamma_{\varsigma\varsigma}^{\mathrm{grav}} = & \frac{1}{16 \pi G_k} \int \mathrm d\t \mathrm d^3\!x\,\sqrt{\sb}\,\tfrac{1}{2} \varsigma \Big[ - \tfrac{2}{9} (\Delta - \bar R ) - \tfrac{4}{3} \Lambda_k - \tfrac{2}{3} \partial_\tau^2 + \gamma \tfrac{12\lambda_k-7}{3} \partial_\tau + \gamma^2\tfrac{11(3\lambda_k - 1)}{6} \Big] \varsigma \nonumber \\ \nonumber\\
\delta^2 \Gamma_{hh}^{\mathrm{grav}} = & \frac{1}{16 \pi G_k} \int \mathrm d\t \mathrm d^3\!x\,\sqrt{\sb}\,\tfrac{1}{2} h \Big[ - \tfrac{2}{9} \Delta + \tfrac{1}{18} \bar R + \tfrac{1}{3} \Lambda_k + \tfrac{3\lambda_k-1}{3} \partial_\tau^2 + \gamma \tfrac{3\lambda_k-1}{6}\partial_\tau + \gamma^2 \tfrac{13(3\lambda_k-1)}{24} \Big] h \nonumber \\ \nonumber\\
\delta^2 \Gamma_{h\varsigma}^{\mathrm{grav}} = & - \frac{1}{16 \pi G_k} \int \mathrm d\t \mathrm d^3\!x\,\sqrt{\sb}\,\tfrac{2}{9} h \, \Big[ \Delta^2 - \tfrac{1}{2} \bar R \Delta \Big] ^{1/2} \, \varsigma \, .
\end{align}
The second variations with respect to the component fields (\ref{2ndvariations3}) are not yet diagonal, but we can handle the mixing term in the scalar sector as shown below. Note as well that $\delta^2 \Gamma_{\xi\xi}^{\mathrm{grav}}$ is independent of the spatial Laplacian, since this will be important in the following. To complete the information about the second variation of the effective average action we give the results for the ghost part. Here it is important that we used the Landau gauge and all fluctuating lapse functions and shift vectors vanish. This simplifies the ghost action (\ref{ghostaction}) considerably and the part quadratic in the ghost fluctuations, once having set background ghosts to zero, reads
\begin{align}\label{ghostvariations}
\delta^2 \Gamma_{\bar{c}c}^{\mathrm{gh}} = & \int \mathrm d\t \mathrm d^3\!x\,\sqrt{\sb}\,\bar c \, \partial_\tau \, c \, , \nonumber \\
\delta^2 \Gamma_{\bar{c}^ic_j}^{\mathrm{gh}} = & \int \mathrm d\t \mathrm d^3\!x\,\sqrt{\sb}\,\bar{c}^i \, \partial_\tau \, c_i \, .
\end{align}
As we have seen for $\delta^2 \Gamma_{\xi\xi}^{\mathrm{grav}}$ we do not get any dependencies on the spatial Laplacian.

Given the variations (\ref{2ndvariations3}) and (\ref{ghostvariations}), we are now in a position to specify the IR
regulators $\R_k$ along the lines of \cite{Rechenberger:2012dt}. We choose a purely spatial regulator, \emph{i.e.} $\R_k=\R_k(\Delta)$ will be a function of the Laplacian on the spatial slice only. In practice we will implement a regulator of Type I, in the nomenclature of \cite{Codello:2008vh}, dressing the spatial Laplacians with a $k$-dependent mass term ($k$ being a three-momentum) according to
\begin{equation}
 \Delta\mapsto\Delta+k^2R^{(0)}(\Delta/k^2)\equiv P_k
\end{equation}
where $R^{(0)}$ is the shape function introduced in (\ref{cutoffstructure}). This choice entails that second variations which are independent of the spatial Laplacian do not get any regulator contributions and therefore drop out of the right hand side of the flow equation (\ref{wettericheqn}). Thus we get no ghost contributions and no vector contributions. We are left with the following regulator
\begin{equation} \label{eq:foliatedRegulator}
\mathcal R_k = \begin{pmatrix} \mathcal R_k^\mathrm{TT} & & \\ & \mathcal R_k^{hh} & \mathcal R_k^{h\varsigma} \\ & \mathcal R_k^{\varsigma h} & \mathcal R_k^{\varsigma\varsigma} \end{pmatrix}
\end{equation}
with
\begin{align}
\R_k^{hh} = \mathcal R_k^{\varsigma\varsigma} &= - \frac{1}{32 \pi G_k} \,\frac{2}{9}\,R_k \, , \nonumber \\
\R_k^{\varsigma h} = \R_k^{h \varsigma} &= - \frac{1}{32 \pi G_k} \, \frac{2}{9} \, \left[ \left(P_k^2 - \tfrac{1}{2} \bar R P_k \right)^{1/2} - \left(\Delta^2 - \tfrac{1}{2} \bar R \Delta \right)^{1/2} \right] \, , \nonumber \\
\R_k^\mathrm{TT} &= \frac{1}{32 \pi G_k} \field{I}_2\,R_k \, .
\end{align}
Here $\field{I}_2$ denotes the two-dimensional identity matrix.

At this stage we have all the ingredients to evaluate the Wetterich equation \eqref{wettericheqn} for the ansatz \eqref{ADMaction}.
Combining the second variations \eqref{2ndvariations3} and the regulator (\ref{eq:foliatedRegulator}) we obtain
\begin{equation} \label{eq:foliatedRHS}
\partial_t\Gamma_k = \mathcal T_\mathrm{TT} + \mathcal T_\mathrm{scalar} \, . 
\end{equation}
The contribution of the transverse traceless modes is collected in $\mathcal T_\mathrm{TT}$ and reads
\begin{align} \label{eq:foliatedTT}
\mathcal T_\mathrm{TT} = 16\pi G_k\, \mathrm{Tr} \bigg[ &\Big( P_k + b_1 + b_2 \bar R + b_3\gamma \p_\t + b_4 \gamma^2 \Big)^{-1}  \, \partial_t\mathcal R_k^\mathrm{TT} \bigg] \; ,
\end{align}
where
\be\label{bdefs}
b_1 = -2\Lambda_k-\p_\t^2 \, , \qquad b_2  = \frac{2}{3} \, , \qquad 
b_3 = \frac{12\lambda_k-7}{2} \, , \qquad b_4 = \frac{9\lambda_k+1}{4} \; .
\ee
The scalar part is a bit more involved since it is given by a $2\times2$ matrix acting on scalar fluctuations. Following the method introduced in \cite{Lauscher:2001ya}, we can treat the scalar part as matrix valued in field space to find
\begin{align} \label{eq:foliatedS}
\mathcal T_\mathrm{scalar} &= \frac{1}{2} \mathrm{Tr} \Big[ \begin{pmatrix} \bar{\Gamma}_{hh}^{(2)} & \bar{\Gamma}_{h\varsigma}^{(2)} \\ \bar{\Gamma}_{\varsigma h}^{(2)} & \bar{\Gamma}_{\varsigma\varsigma}^{(2)} \end{pmatrix}^{-1} \partial_t \begin{pmatrix} \mathcal R_k^{hh} & \mathcal R_k^{h\varsigma} \\ \mathcal R_k^{\varsigma h} & \mathcal R_k^{\varsigma\varsigma} \end{pmatrix} \Big] \nonumber \\
&= \frac{1}{2} \mathrm{Tr} \Big[ \frac{\bar{\Gamma}_{hh}^{(2)}\partial_t\mathcal R_k^{\varsigma\varsigma} + \bar{\Gamma}_{\varsigma\varsigma}^{(2)}\partial_t\mathcal R_k^{hh} - 2 \bar{\Gamma}_{h\varsigma}^{(2)}\partial_t\mathcal R_k^{h\varsigma}}{\bar{\Gamma}_{hh}^{(2)} \bar{\Gamma}_{\varsigma\varsigma}^{(2)} - \bar{\Gamma}_{h\varsigma}^{(2)} \bar{\Gamma}_{\varsigma h}^{(2)}} \Big]
\end{align}
where $\mathrm{Tr}$ indicates the integral over the four-dimensional momentum space and $\left[\bar{\Gamma}^{(2)}\right]_{ij} \equiv \left[\Gamma^{(2)} + \mathcal R_k\right]_{ij}$.

\subsection{Evaluating the functional traces}
\label{sect3.2}
%
The beta functions capturing the scale dependence of the coupling constants of our truncation can be constructed as follows. The flow equation for the Newton's constant $G_k$ can be found by projections onto the terms proportional to $\bar R$. The one for the cosmological constant $\Lambda_k$ follows from the volume term and finally the flow equation for the anisotropy coupling $\lambda_k$ can be found by projection onto terms proportional to $\gamma^2$. Higher orders in $\bar R$ and $\gamma$ do not contribute to the running of these couplings, and are therefore not of interest in the present computation. This can be used to evaluate the right hand side of the Wetterich equation within our truncation (\ref{eq:foliatedRHS}). In this subsection, we carry out this computation for the transverse traceless part. The scalar sector can be treated in a similar way and the explicit formulas are given in Appendix \ref{app:scalarTrace}. The expansion of \eqref{eq:foliatedTT} in $\bar R$ and $\gamma$ leads to
\be\label{expandedTT}
\mathcal T_\mathrm{TT} =  16\pi G_k\, \mathrm{Tr}\,\bigg[ \, \partial_t\mathcal R_k^\mathrm{TT} \; \Big( \frac{1}{P_k+b_1} - \frac{b_2}{(P_k+b_1)}\bar R 
- \frac{b_4}{(P_k+b_1)^2}\gamma^2 + \frac{b_3^2}{(P_k+b_1)^3}\gamma^2 \p_\t^2 \Big) \bigg] + \ldots \; .
\ee
Here and in the following, the dots denote irrelevant terms which do not carry information about the running 
of the coupling constants. Note that in particular the term linear in $\gamma$ does not contribute to
the truncation.

Up to this point we regulated only the spatial fluctuations but ignored the fluctuations in time direction. 
These will be treated along the lines of finite-temperature flow equations \cite{Litim:1998yn,Litim:2006ag,Floerchinger:2011sc} by introducing a circle of finite length $T$ in the $\tau$ direction and impose periodic boundary conditions. This can be used to Fourier expand the $\t$-dependent fields with a discrete Fourier transformation
\begin{equation}
\phi(\tau,x) = \!\! \sum_{n=-\infty}^\infty \!\! \phi_n(x) e^{2 \pi \imath n \tau /T} \quad \Rightarrow \quad \phi_n(x) = \frac{1}{T}\int_0^T \!\!\! d\tau \, \phi(\tau,x)e^{-2 \pi \imath n \tau /T} \; .
\end{equation}
The complex coefficients $\phi_n(x)$ satisfy the reality constraint $\phi_n(x)=\phi_{-n}^\ast(x)$. In this case 
the $\t$ derivatives are converted into Matsubara masses of the fluctuations in time direction
\begin{equation}\label{derivexpansion}
 \p_\t\,\to\,\imath\,\frac{2\pi n}{T}\;.
\end{equation}

To further analyze the Fourier expansion of our truncation, it is convenient to further specify the (up to now general) warp function $\c$
by setting
\begin{equation}
 \c(\t)=e^{-\t/\q} \, . 
\end{equation}
Here $\theta$ should be considered as a geometrical parameter that can be used to
distinguish the terms carrying the relevant information for reading off the 
beta functions. Indeed, substituting this ansatz into \eqref{prolhs} we find
that the geometric invariants encoding the scale dependence of $\lambda_k, \Lambda_k$ and
$G_k$ have the distinguished signatures
\begin{equation}\label{pro2lhs}
\begin{split}
 \int_0^T\!\!\! \mathrm d\tau \, \sqrt{\sb}\,=\;&\int_0^T\!\!\! \mathrm d\tau \, \c^{3/2}\sqrt{\bar{s}}\,\to\, T(\sqrt{\bar{s}}+\O(T /\q)) \, ,\\
 \int_0^T\!\!\! \mathrm d\tau \, \sqrt{\sb}\,\bar{R}\,=\;&\int_0^T\!\!\! \mathrm d\tau \, \c^{1/2}\sqrt{\bar{s}}\;\SR\,\to\,T\,\SR\,(\sqrt{\bar{s}}+\O(T /\q)) \, ,\\
 \int_0^T\!\!\! \mathrm d\tau \, \sqrt{\sb}\,\g^2\,=\;&\int_0^T\!\!\! \mathrm d\tau \, \c^{3/2}\sqrt{\bar{s}}\,/\,\q^2\,\to\,T/\,\q^2(\sqrt{\bar{s}}\,+\O(T /\q))\;.
\end{split}
\end{equation}
Substituting \eqref{pro2lhs} into \eqref{prolhs} and subsequently applying the $t$ derivative, the left-hand side of the flow equation to lowest order in $T/\q$ becomes
\begin{align}\label{LHS}
 \p_t{\Gamma}_k
 &=\frac{1}{16\pi}\int \mathrm d^3\!x\,\sqrt{\bar{s}}\,\left[\frac{3}{4\q^2}\left(\p_t\frac{T}{G_k}-3\,\p_t\frac{T\l_k}{G_k}\right)-\,\SR\,\p_t\frac{T}{G_k}+2\,\p_t\frac{T\Lambda_k}{G_k}\right] \, .
\end{align}
Performing the time integral led to a factor $T$ that has been kept inside the $t$ derivatives to take into account a possible scale dependence of $T = T(k)$, a possibility that will be investigated later on. This expression shows that all the required information
on the beta functions can be obtained at zeroth order in the parameter $T/\q$.
This observation can be used to crucially simplify the Fourier analysis of the
right-hand side, since it allows to set all functions $\chi$ which do not come with a time derivative equal to one. Strictly speaking, this substitution should be considered
as a projection of the right-hand side, stating that only the constant Fourier mode of $\chi$ contributes in the projection of the traces. 

The discussion about the right-hand side is slightly more complex. The four-dimensional
trace $\mathrm{Tr}$ decomposes into a sum over Fourier-modes in $\tau$ and a trace $\mathrm{tr}$
defined on a spatial slice. Schematically, 
\begin{equation}
 \mathrm{Tr}\left[ \cO \right] = \sum_{n=-\infty}^\infty \, \mathrm{tr} \, \left[ \cO \right]_n \;.
\end{equation}
Applying this relation to the first term of the transverse traceless part (\ref{expandedTT})
yields 
\be
\mathrm{Tr}\left[\frac{\partial_t(T\,Z_{\mathrm Nk}R_k)}{T\,Z_{\mathrm Nk}}\,\frac{1}{P_k+b_1}\right]
= \sum_n \mathrm{tr}\left[ 
\frac{\partial_t(T\,Z_{\mathrm Nk}R_k)}{T\,Z_{\mathrm Nk}} \left(\frac{1}{P_k-2\Lambda_k+\left(\tfrac{2\pi n}{T}\right)^2 }\right)\,
\right] +\O(T/\q) \;.
\ee
Here we introduced the wave function renormalization $Z_{\mathrm Nk}$, defined by $G_k = Z_{\mathrm Nk}^{-1} G_0$, carrying the scale dependence  
of Newton's constant. The other terms appearing
in $\mathcal T_{\mathrm{TT}}$ follow along the same lines.

The final step in evaluating (\ref{expandedTT}) consists in performing 
the $\rm tr$ and summing the Matsubara modes. This can be done
by applying the trace technology \cite{Rechenberger:2012dt}.\footnote{
In this course, we also corrected a misprint in the heat kernel coefficients of the constrained fields used in \cite{Rechenberger:2012dt},
which did not influence the results \cite{SRPhD}.} The result is most conveniently expressed in terms of the dimensionless analogs of Newton's constant, the cosmological constant and the Matsubara mass $m$
\begin{equation}\label{dimlessqs}
 \Gt_k=G_k\,k^2\quad,\qquad\Lt_k=\Lambda_k\,k^{-2}\quad,\qquad m=\frac{2\pi}{Tk} \, . 
\end{equation}
Performing the tr using standard heat-kernel techniques in combination with the optimized cutoff $R_k(z)=k^2 (1-z) \,\q(1-z)$ \cite{opt}, yields
\begin{align} \label{eq:TTT}
\mathcal T_\mathrm{TT} = &\frac{2k^3}{(4 \pi)^{3/2}} \sum_n \int \mathrm d^3\!x\,\sqrt{\bar{s}}\, \Big[ q^{1,0}_{3/2} - \tfrac{\bar R}{k^2} \left( \tfrac{5}{6}\, q^{1,0}_{1/2} + b_2 \, q^{2,0}_{3/2} \right) 
- \tfrac{\gamma^2}{k^2}\Big( b_3^2 \, q_{3/2}^{3,0} + b_4 \, q_{3/2}^{2,0} \Big) \Big] + \ldots \, .
\end{align}
Here the threshold functions $q^{p,q}_n(w) \equiv \Phi^{p,q}_n(w) - \tfrac{1}{2} \eta_N \tilde{\Phi}^{p,q}_n(w)$ are constructed from 
\begin{equation}\label{eq:optimisedThreshold}
 \Phi^{p,q}_n(w)=\frac{1}{\Gamma(n+1)}\,\frac{1}{(1+w)^p}\;,\qquad\tilde{\Phi}^{p,q}_n(w)=\frac{1}{\Gamma(n+2)}\,\frac{1}{(1+w)^p}
\end{equation}
and evaluated at 
\begin{equation}\label{eq:w2T}
w_\mathrm{TT} = -2\Lt_k + m^2n^2\;.
\end{equation}
The  anomalous dimension of the Newton's constant is defined as
\be\label{etaN}
 \eta_\mathrm N \equiv -\partial_t \ln(T\,Z_{\mathrm Nk}).
\ee
This definition of the anomalous dimension differs from the usual one in the sense that it also includes the (potentially scale dependent) $T$.
Finally, the sum over Matsubara modes can be carried out analytically. Following Appendix \ref{app:MatsubaraSum} the result can be given in terms of the summed threshold functions $T^{p,q}_l$ defined in \eqref{eq:summedq2T} and reads
\begin{align} \label{eq:summedRHS}
\mathcal T_\mathrm{TT} = &\frac{2k^3}{(4 \pi)^{3/2}} \int \mathrm d^3\!x\,\sqrt{\bar{s}}\, \Big[ T^{1,0}_{3/2} - \tfrac{\bar R}{k^2} \left( \tfrac{5}{6} T^{1,0}_{1/2} + b_2 \, T^{2,0}_{3/2} \right)
- \tfrac{\gamma^2}{k^2}\Big( b_3^2 \, T_{3/2}^{3,1} + b_4 \,  T_{3/2}^{2,0} \Big) \Big] \, . 
\end{align}

The evaluation of the scalar trace follows along the same lines and the corresponding details can be found in Appendices \ref{app:scalarTrace} and \ref{app:MatsubaraSum}. Here we only state the final result expressed in terms of the functions $S$ defined in \eqref{eq:summedqScalar} and the expansion coefficients $a_i$ given in \eqref{eq:foliatedScalarPrefactors}
\begin{align} \label{eq:T0}
\mathcal T_\mathrm{scalar} = &-\frac{2k^3}{(4\pi)^{3/2}} \int \mathrm d^3\!x\,\sqrt{\bar{s}}\, \bigg[ a_1 S_{3/2,0}^{1,0} 
+ \tfrac{\bar R}{k^2} \Big( \tfrac{a_1}{6}S_{1/2,0}^{1,0} + a_2S_{3/2,1}^{1,0} + (a_3+a_5)S_{3/2,1}^{2,0} + a_4 S_{3/2,1}^{2,1} \Big) \nonumber \\
& + \tfrac{\gamma^2}{k^2} \Big( a_6S_{3/2,1}^{1,0} + (a_7+a_{11})S_{3/2,1}^{2,0} + a_8S_{3/2,1}^{2,1} + (a_9+a_{12})S_{3/2,2}^{2,1} \nonumber \\
& \qquad\quad + a_{10}S_{3/2,2}^{2,2} + (a_{13}+a_{16}+a_{18})S_{3/2,2}^{3,1} 
 + (a_{14}+a_{17})S_{3/2,2}^{3,2} + a_{15}S_{3/2,2}^{3,3} \Big) \bigg] \; .
\end{align}
The expressions \eqref{eq:summedRHS} and \eqref{eq:T0} complete the evaluation of the functional traces appearing in our truncation.

The final form of the beta functions in terms of the functions $T$ and $S$ are found by comparing the coefficients
multiplying the geometric monomials $\gamma^0 \bar{R}^0$, $\gamma^0 \bar{R}$ and $\gamma^2 \bar{R}^0$ on the left-
and right-hand side. Expressed in terms of the dimensionless coupling constants \eqref{dimlessqs} these become
\begin{align} \label{eq:folFinalSystem}
\partial_t \Gt_k = \beta_{\Gt}(\Gt_k, \Lt_k, \lambda_k, m) \; , \nonumber \\
\partial_t \Lt_k = \beta_{\Lt}(\Gt_k, \Lt_k, \lambda_k, m) \; , \nonumber \\
\partial_t \lambda_k = \beta_\lambda(\Gt_k, \Lt_k, \lambda_k, m) \; ,
\end{align}
with the explicit form of the beta functions given by
\begin{align}\label{folFinalBetas}
\beta_{\Gt} =& (2+ \eta_\mathrm N + \tfrac{\partial_t T}{T})\Gt_k \; , \nonumber \\
\beta_{\Lt} =& (\eta_\mathrm N - 2)\Lt_k + \frac{4m\Gt_k}{(4\pi)^{3/2}}\left[ 2T^{1,0}_{3/2} - 2a_1S^{1,0}_{3/2,0} \right] \; , \nonumber \\
\beta_\lambda =& \frac{3\lambda_k-1}{3}\eta_\mathrm N +\frac{64m\Gt_k}{9(4\pi)^{3/2}} \Big[ \tfrac{(12\lambda_k-7)^2}{4}T_{3/2}^{3,1} + \tfrac{9\lambda_k+1}{4} T_{3/2}^{2,0} + a_6S_{3/2,1}^{1,0} + (a_7+a_{11})S_{3/2,1}^{2,0} + a_8S_{3/2,1}^{2,1} \nonumber \\
& + (a_9+a_{12})S_{3/2,2}^{2,1} + a_{10}S_{3/2,2}^{2,2} + (a_{13}+a_{16}+a_{18})S_{3/2,2}^{3,1} + (a_{14}+a_{17})S_{3/2,2}^{3,2} + a_{15}S_{3/2,2}^{3,3} \Big] \; .
\end{align}
The anomalous dimension \eqref{etaN} is obtained from the coefficients multiplying the $\bar{R}$ terms, yielding
\be
\eta_\mathrm N = -\frac{16m\Gt_k}{(4\pi)^{3/2}}\Big[ \tfrac{5}{6}T_{1/2}^{1,0} + \tfrac{2}{3}T_{3/2}^{2,0} + \tfrac{a_1}{6}S_{1/2,0}^{1,0} + a_2S_{3/2,1}^{1,0} + (a_3+a_5)S_{3/2,1}^{2,0} + a_4 S_{3/2,1}^{2,1} \Big] \; .
\ee
Taking into account that the functions $T$ and $S$ also carry an implicit $\eta_{\rm N}$ dependence, this equation can be solved for $\eta_{\rm N}$
and written in terms of the functions $\Psi$, $\tilde\Psi$, $\Upsilon$ and $\tilde\Upsilon$ given explicitly in Appendix \ref{app:MatsubaraSum}
\begin{equation} \label{eq:HLanoDim}
\eta_\mathrm N = \frac{m\Gt_k B_1(\Lt_k, \lambda_k)}{1+m\Gt_k B_2(\Lt_k, \lambda_k)}
\end{equation}
with
\begin{align}
B_1 =& -\frac{16}{(4\pi)^{3/2}}\Big[ \tfrac{a_1}{6}\Psi_{1/2,0}^{1,0} + a_2\Psi_{3/2,1}^{1,0} + (a_3+a_5)\Psi_{3/2,1}^{2,0} + a_4 \Psi_{3/2,1}^{2,1} + \tfrac{5}{6}\Upsilon_{1/2}^{1,0} + \tfrac{2}{3}\Upsilon_{3/2}^{2,0} \Big] \nonumber \\
B_2 =& -\frac{8}{(4\pi)^{3/2}}\Big[ \tfrac{a_1}{6}\tilde\Psi_{1/2,0}^{1,0} + a_2\tilde\Psi_{3/2,1}^{1,0} + (a_3+a_5)\tilde\Psi_{3/2,1}^{2,0} + a_4 \tilde\Psi_{3/2,1}^{2,1} + \tfrac{5}{6}\tilde\Upsilon_{1/2}^{1,0} + \tfrac{2}{3}\tilde\Upsilon_{3/2}^{2,0} \Big] \; .
\end{align}
The beta functions \eqref{folFinalBetas} and the anomalous dimension \eqref{eq:HLanoDim} constitute the main results of this section and provide 
the starting point for the systematic study of the RG flow of projectable HL gravity at low energies in the next section.
Notably, the beta functions depend parametrically on the Matsubara mass $m$. This reflects the feature
that the anisotropic quantum field theory possesses \emph{two} correlation lengths associated with the time-
and spatial sector of the theory, respectively. 

\section{RG flow of anisotropic gravity models I}
\label{sec:RGflow}
In this section we analyze the properties of the beta functions \eqref{folFinalBetas} for the case that the dimensionless Matsubara mass $m$ is fixed to a constant value.
\subsection{Running $T$: the floating fixed point scenario} \label{subsec:floating}
As already observed in Section \ref{sect3.2}, the beta functions \eqref{folFinalBetas} parametrically depend on the Matsubara mass $m$.
In order to complete the system, one has to make an assumption on the scale dependence of $m$. Technically, this 
information should be provided by a beta function controlling the flow, $\p_t m_k = \beta_m$.
The computation of $\b_m$ is beyond the scope of the present work, however. Therefore, we will follow \cite{Manrique:2011jc}, and first adopt the ``floating fixed point scenario'', assuming that the periodicity $T$ is inversely proportional to the RG scale, $T\propto k^{-1}$. This choice essentially corresponds to specifying the anisotropic regulator of the system. Setting $T\propto k^{-1}$
thereby imposes that fluctuations in space and time direction are regulated by the same $k$ dependence.\footnote{It would be interesting to study the effect of imposing different regulators in the flow equation in a systematic way. We hope to come back to this point in the future.} This choice entails that $m$ is actually independent of $k$, fixing
\begin{equation}
 \b_m=0\;.
\end{equation}
As a result, the Matsubara mass is assumed to have a non-trivial fixed point $m_\ast\neq0$. 
A direct consequence of such choice is that the time circle automatically decompactifies as the theory flows towards the IR, as one would expect from a physically viable theory. On the other hand in the deep UV the circle collapses in a controlled way, so that the presence of the ``time direction'' still enters into the beta functions through the dimensionless parameter $m_\ast$.
We then adopt $m_\ast$ as a free parameter and study the parametric dependence of (\ref{eq:folFinalSystem}) on it.

\subsection{Analytic structure of the beta functions}
We start with investigating the analytic structure of the threshold functions $T$ and $S$ defined in \eqref{eq:summedq2T} and \eqref{eq:summedqScalar}.
As already pointed out in \cite{Rechenberger:2012dt}, these are build from hyperbolic and trigonometric functions or a mixture of the two. This analytic
structure turns out to be independent of $\tilde G$ but actually depends on the values of $\Lt$ and $\l$. While the hyperbolic functions are well defined over the whole space, the trigonometric terms give rise to a rich pole structure in the beta functions. The origin of these poles can be traced back to the wrong kinetic sign of the conformal factor $h$ plaguing not only the Einstein-Hilbert action but also appearing in the ansatz \eqref{ADMaction}. Presumably, this problem can be cured by including higher-derivative terms which are presently not contained in the truncation.
\begin{figure}[!t]
\begin{center}
 \includegraphics[width=12cm,height=6cm]{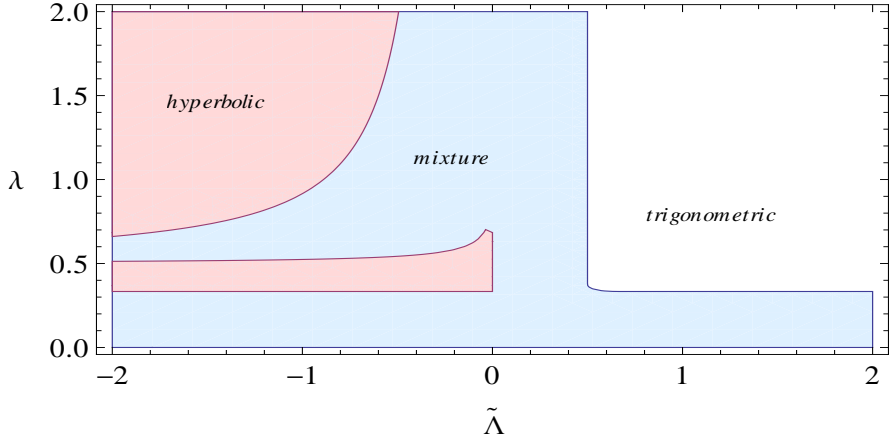}
\end{center}
\caption{The analytic structure of the beta functions (\ref{folFinalBetas}) as functions of the cosmological constant $\Lt$ and the anisotropy parameter $\l$, evaluated at $m_\ast=1$. We distinguish the cases where the threshold functions are build from hyperbolic or trigonometric functions only or contain both types of terms.}
\label{fig:structure}
\end{figure}

Figure \ref{fig:structure} shows the distribution of hyperbolic, mixture and trigonometric areas in the space of couplings. Although it represents an extremely important piece of information about the regions of the truncated theory space that are accessible to the flow, the presence of trigonometric functions does not \emph{a priori} indicate the inaccessibility of a region unless we are taking the \emph{decompactification} limit $m_\ast\to0$. In this case, as already discussed in \cite{Rechenberger:2012dt}, the discrete Matsubara sums become continuous integrals. As a consequence, the trigonometric terms in the beta functions develop a branch-cut singularity leading to diverging expressions. Thus, in this limit, the beta functions are only well defined where they consist of hyperbolic terms.

\begin{figure}[p]
\begin{center} 
 \includegraphics[width=12cm,height=6cm]{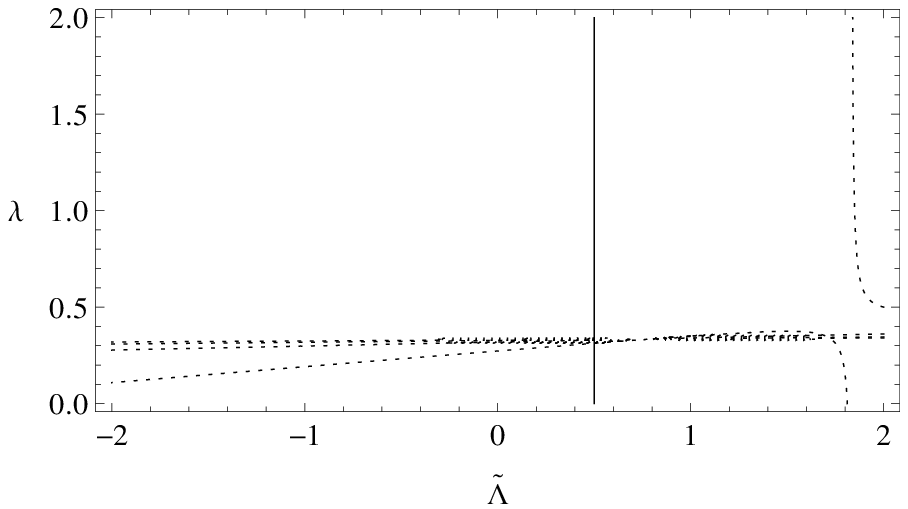}\\
 \includegraphics[width=12cm,height=6cm]{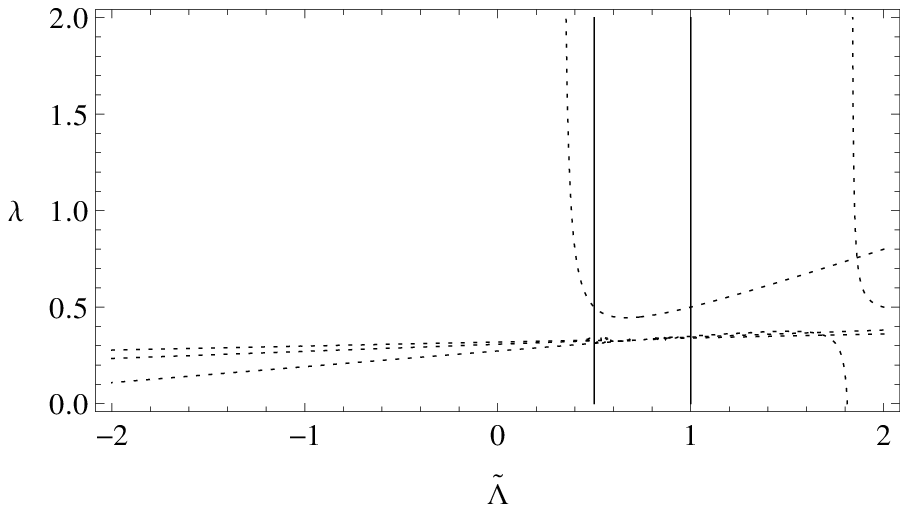}\\
 \includegraphics[width=12cm,height=6cm]{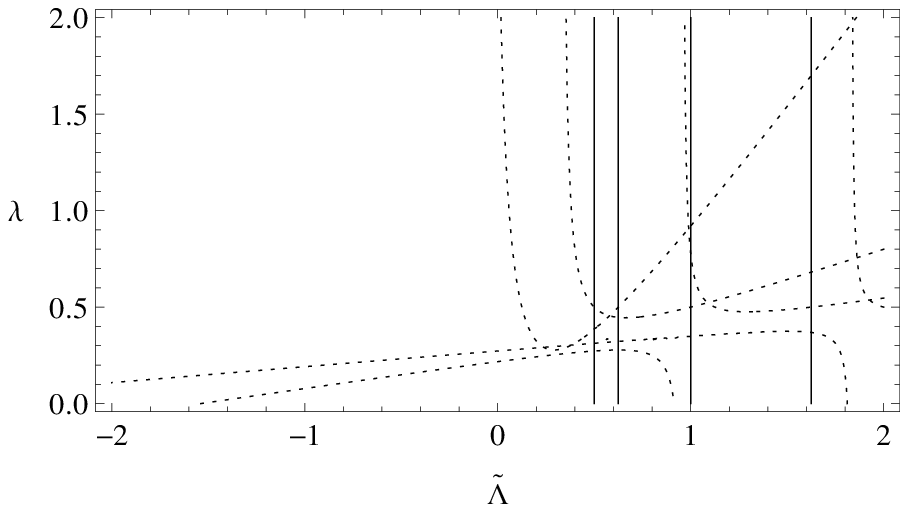}\\
\end{center}
\caption{The singularity structure of the beta functions induced by the trigonometric functions, evaluated at $m_\ast=2$ (upper panel), $m_\ast=1$ (central panel) and $m_\ast=1/2$ (lower panel). Solid lines indicate the poles generated by the transverse-traceless part of the right-hand side, dashed lines by the scalar part. The emergence of new singular loci for decreasing Matsubara masses is clearly visible.}
\label{fig:poles}
\end{figure}

In the case of a finite Matsubara mass, we will need to explicitly derive the pole structures generated by the trigonometric functions. The poles 
appear whenever one of the trigonometric functions appearing in the denominators of $T$ and $S$ vanishes. Since the arguments
of the trigonometric functions are independent of Newton's constant, these singular loci can be visualized as curves on the $\{\Lt,\l\}$-plane.
These structures are shown in Figure \ref{fig:poles} for several values of $m_\ast$.
It is easy to notice that the singularity-free areas at the top left of the diagrams in Figure \ref{fig:poles} are considerably larger than the hyperbolic region shown in Figure \ref{fig:structure}. They also include a part of the mixture region. Besides, the accessible region with $\Lt\leq1/2$, that is a common feature of the metric calculations \cite{Reuter:2001ag}, is hardly affected by the additional structures found in this analysis, at least for $m_\ast\gtrsim1$. For lower values, a proliferation of singularities
shrinks the region where the beta functions are well defined, see Figure \ref{fig:poles}. The low energy regime of the theory, that is expected (and will be shown in the following) to be located in the area around
\begin{equation}
 \Gt\gtrsim0\quad,\qquad \Lt\simeq 0 \quad,\qquad \l\simeq1\;,
\end{equation}
will also be affected by these singularities if $m_\ast \lesssim 1$. Thus we restrict ourselves in the sequel to $m_\ast\geq1$, in order to 
avoid the proliferation of singularities entailed by small values $m_\ast$.

\subsection{Fixed points of the beta functions}
\label{finitem}
A crucial property of the beta functions \eqref{eq:folFinalSystem} are their fixed points, defined as points $\tilde{g}_{\ast,i} \equiv \{\Gt_\ast,\Lt_\ast,\l_\ast\}$ in the space of the couplings that exhibit a simultaneous vanishing of the whole system
\begin{equation}\label{fpcond}
 \b_{\tilde{g}_i}(\tilde{g}_{\ast,i})=0 \, .
\end{equation}
Their importance can hardly be overstated. Firstly, they may provide a consistent and predictive UV-completion for RG trajectories being
attracted to the fixed point at high energies. This can either be in the context of asymptotic freedom (Gaussian Fixed Point) 
or Asymptotic Safety (non-Gaussian Fixed Point). Moreover, for RG trajectories passing sufficiently close, the fixed point
can imprint a specific scaling behavior on the coupling constants. Provided that the RG trajectory spends sufficient RG time
in its vicinity, the fixed point can induce scaling regimes where physics is governed by the same scaling for many orders of magnitude. 
The information on possible scaling behaviors is encoded in the  
the stability matrix 
\begin{equation}\label{stab.mat}
 M_{ij}=\left.\frac{\partial\b_{\gt_i}}{\partial\gt_j}\right|_{\gt_\ast}\;.
\end{equation}
 The critical exponents $\q_i$, defined as minus the eigenvalues of $M$, carry the desired information: a positive (negative) critical exponent denotes an UV-attractive (UV-repulsive) scaling direction. The directions are specified by the eigenvectors $v_i$ associated with the corresponding eigenvalue $\theta_i$,
\begin{equation}
 Mv_i = -\q_i \,v_i \;.
\end{equation}

Searching for fixed points of the system \eqref{eq:folFinalSystem}, one first encounters a \emph{one-parameter family} of Gaussian Fixed Points (GFPs). Linearizing 
the system at $\Gt_\ast=0$, $\Lt_\ast=0$ yields
\begin{align}\nonumber\label{floatlinsys}
 \b_\Gt&\simeq\Gt\\
 \b_\Lt&\simeq\frac{2\Gt}{3\pi}\left(\tfrac{2}{\tanh(\pi/m_\ast)}+\tfrac{m_\ast\sqrt{w}}{\tanh(\pi \sqrt{w})}\right)-2\Lt\nonumber\\
 \b_\l&\simeq \frac{\Gt\, m}{27 \, \pi} \bigg[ 
\tfrac{(7 - 6 \lambda) (13 - 24 \lambda)}{m_\ast \tanh(\pi/m_\ast)}
+ \tfrac{\pi \, (61 - 156 \lambda + 144 \lambda^2) }{m_\ast^2 \sinh(\pi/m_\ast)^2} 
- \tfrac{2 \pi^2 (7 - 12 \lambda)^2}{m_\ast^3 \tanh(\pi/m_\ast) \sinh(\pi/m_\ast)^2}\nonumber\\ 
& \qquad \quad\quad\!\! - \tfrac{1-3\lambda}{(1-\lambda)^3} \Big(
+ \tfrac{\sqrt{w} \, (19 - 204 \lambda + 527 \lambda^2 - 
    534 \lambda^3 + 144 \lambda^4)}{2 \tanh(\pi \sqrt{w}) }\nonumber\\
& \qquad \quad\quad\!\! + \tfrac{\pi \, w \, (13 - 186 \lambda + 509 \lambda^2 - 
   528 \lambda^3 + 144 \lambda^4)}{2 \sinh(\pi \sqrt{w})^2} 
   - \tfrac{\pi^2 \, w^{3/2} \, (3 - 11 \lambda + 
   12 \lambda^2)^2 }{\tanh(\pi \sqrt{w}) \sinh(\pi \sqrt{w})^2} \Big)\nonumber\\
& \qquad \quad\quad\!\! - \tfrac{4(56 - 309 \lambda + 677 \lambda^2 - 753 \lambda^3 + 
    459 \lambda^4 - 162 \lambda^5)}{\pi \, (1-\lambda)^3}
\bigg]\, . 
\end{align}
Here we used that \eqref{eq:wpm} gives rise to the limits
\begin{equation}
 w= - \lim_{\Lt \rightarrow 0} w_+ = \frac{1-\l}{m_\ast^2\,(3\l-1)}\;,
\end{equation}
and $\lim_{\Lt \rightarrow 0} w_- = 0$. 
In contrast to $\b_\Gt$ and $\b_\Lt$, $\b_\l$ does not contain a term linear in only $\lambda$. This is
actually a consequence of $\lambda$ being dimensionless.
Thus setting $\Gt = 0$, $\Lt = 0$ suffices to solve all three conditions \eqref{fpcond}, indicating the existence 
of a line of GFPs parameterized by $\lambda_\ast$:
\be\label{GFP1}
{\rm GFP:} \qquad \quad  \Gt_\ast=0\;,\quad\Lt_\ast=0\;,\quad \lambda = \l_\ast \, . 
\ee 
 The
degeneracy in $\lambda_\ast$ is associated with the freedom of constructing the kinetic term of the free theory from the Wheeler-de Witt metric,
where $\lambda$ appears as a free parameter. 

The line of GFPs \eqref{GFP1} contains the special points $\lambda_\ast^{\rm conf} = 1/3$ and $\lambda_\ast^{\rm EH} = 1$.
For the first value the kinetic term of the conformal fluctuations $h$ decouples (cf.\ \eqref{2ndvariations3}) while for the second value
the action \eqref{ADMaction} actually coincides with the Einstein-Hilbert action. Analyzing the system \eqref{floatlinsys} at $\lambda_\ast = \lambda_\ast^{\rm conf}$
while leaving $\Gt$ finite actually shows that almost all scalar contributions to $\beta_\lambda$ vanish. Thus in this case
the flow of $\lambda$ is driven almost exclusively by the contribution of the transverse-traceless
 part of the metric fluctuations. For generic values $\Gt$ and $m_\ast$ the $\lambda = 1/3$-plane has no particular properties
 with respect to the RG flow.
 
 For the classical Einstein-Hilbert case $\lambda_\ast^{\rm EH} = 1$ the system \eqref{floatlinsys} apparently becomes singular. Expanding in $\lambda = 1$, however,
 we find that the beta functions are completely regular at this point\footnote{The precise cancellation of the poles in $(1-\lambda)$ provides a highly non-trivial check of the computation.}
\begin{align}\nonumber\label{msys1}
 \b_\Gt&\simeq\Gt\\
 \b_\Lt&\simeq\frac{2}{3\pi}\left(m_\ast+\tfrac{2}{\tanh(\pi/m_\ast)}\right)\Gt-2\Lt\\
 \b_\l&\simeq-\frac{1}{27}\left(\tfrac{154m_\ast}{\pi^2}+\tfrac{68\pi^2}{45m_\ast^3}+\tfrac{32\pi^4}{945m_\ast^5}+\tfrac{11}{\pi\tanh(\pi/m_\ast)}-\tfrac{49}{m_\ast\sinh(\pi/m_\ast)^2}-\tfrac{50\pi}{m_\ast^2\tanh(\pi/m_\ast)\sinh(\pi/m_\ast)^2}\right)\Gt\nonumber\;.
\end{align}
This result establishes that also the $\lambda=1$ plane has no distinguished properties in terms of RG flows. Using that for any positive value $m_\ast$ the bracket appearing in $\b_\l$ is
positive, shows that the RG flow will generically just pierce this plane from below and carry on increasing the value of $\lambda$ towards the IR.
Note, however, that for $\Gt \ll 1$, close the this GFP also the running of $\lambda$ will be tiny.

Furthermore,
$\b_\Gt$ implies that $\Gt(k)$ scales as $k$. As a consequence the dimensionful Newton's constant
$G_k = k^{-2} \Gt$ is proportional to $k^{-1}$. Thus $G_k$ increases towards the IR and diverges as $k \rightarrow 0$.
Obviously, this behavior contradicts the expectation that the dimensionful Newton's constant actually freezes out and becomes
constant at terrestrial scales. This rather unusual scaling behavior of $G_k$ close to the GFP has its origin
in the ``floating fixed point scenario''. The assumption $m_k = m_\ast$ entails that the dimensionful scale $T$ does not freeze
out in the IR and remains scale dependent also at the GFP. Thus the ``floating fixed point scenario'' is not suitable
for describing the physics of a GFP where all dimensionless couplings scale according to the inverse mass dimension
of their dimensionful counterparts. This point will be discussed further in the context of the ``interpolating scenario''
advocated in Section 5.

\begin{figure}[!t]
\begin{center}
 \includegraphics[width=5cm,height=3.5cm]{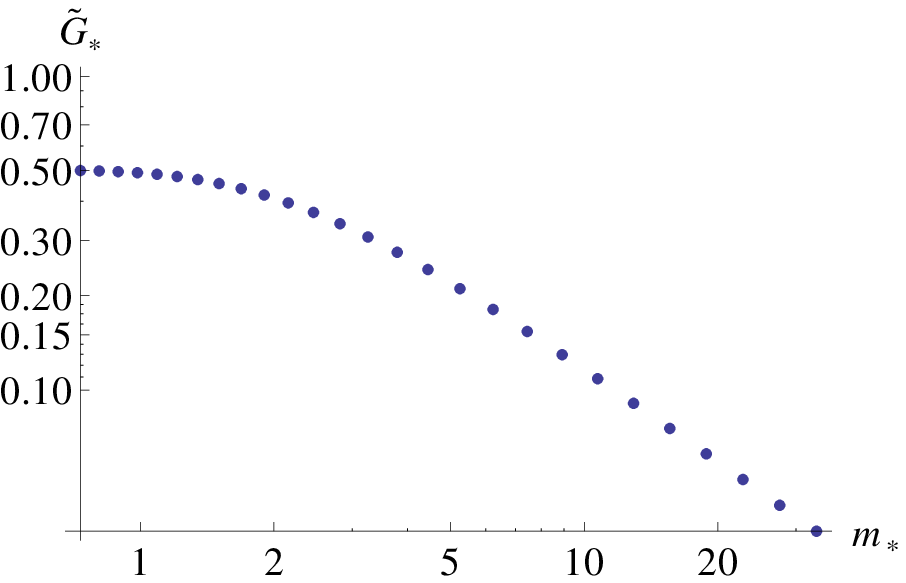}
 \includegraphics[width=5cm,height=3.5cm]{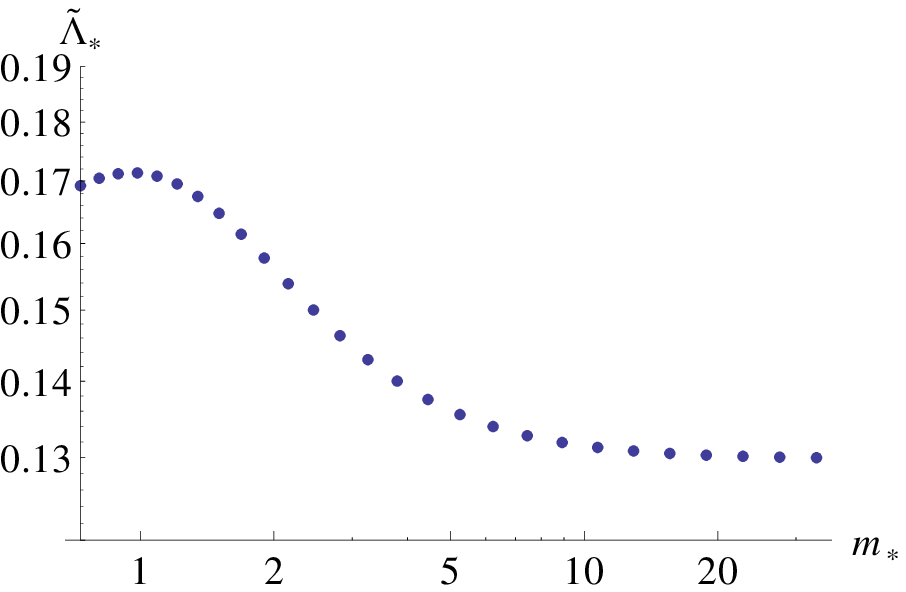}
 \includegraphics[width=5cm,height=3.5cm]{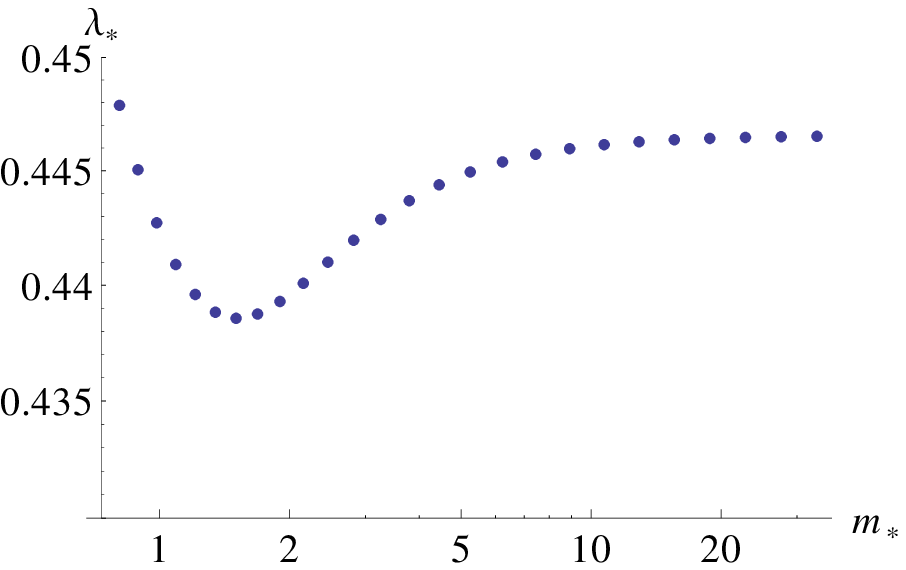}\\
 \includegraphics[width=5cm,height=3.5cm]{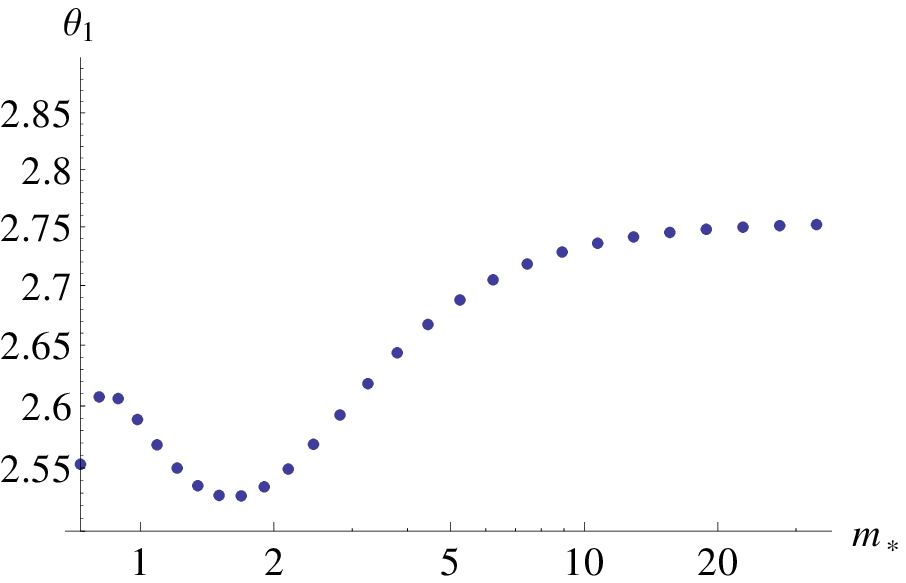}
 \includegraphics[width=5cm,height=3.5cm]{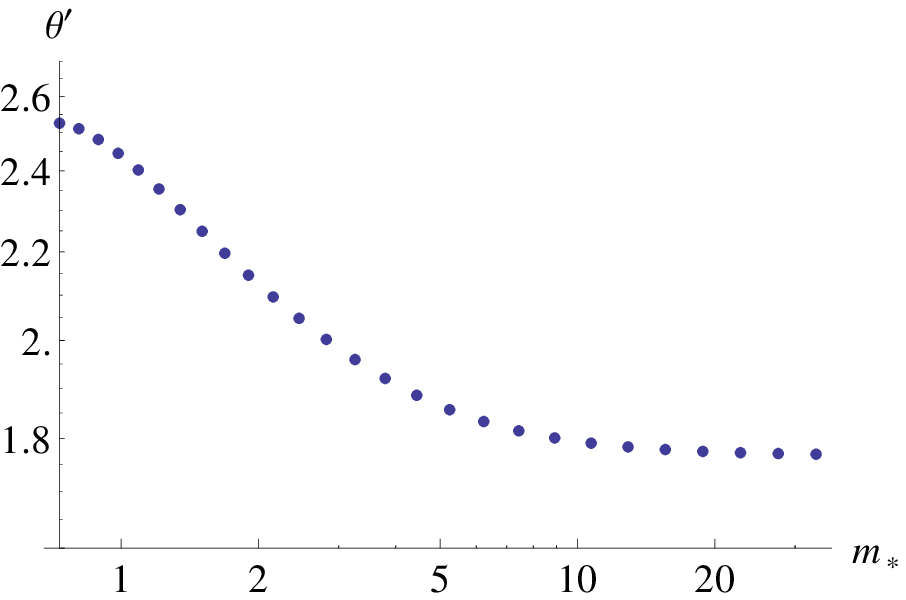}
 \includegraphics[width=5cm,height=3.5cm]{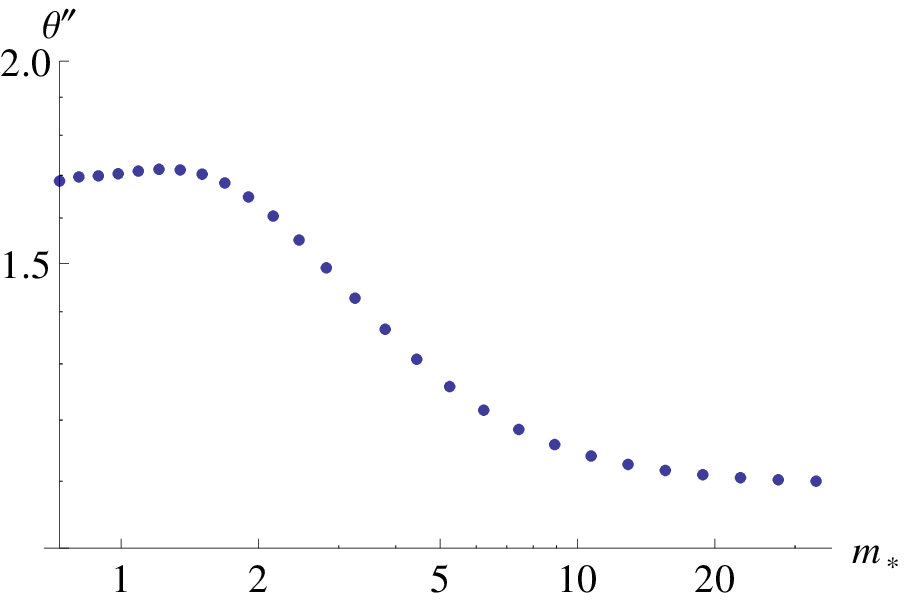}
\end{center}
\caption{
Parametric $m_\ast$ dependence of the position of the NGFP (upper panels) and its stability coefficients $\q_1$ and $\q_\pm=\q'\pm\imath\,\q''$ (lower panels). For $m_\ast \to \infty$ the
location of the NGFP approaches \eqref{ngfp1}. For small Matsubara mass $m_\ast \ll 1$ the properties of the NGFP are affected by the proliferation of singularities in the beta functions.
}
\label{fig:ngfps}
\end{figure}

A numerical investigation reveals that the system \eqref{eq:folFinalSystem}  also possesses a non-Gaussian fixed point (NGFP).
Its precise location and critical exponents depend on the value of $m_\ast$. The first line of diagrams displayed in Figure
\ref{fig:ngfps} depicts the position of this NGFP for a wide range of values of the Matsubara-mass parameter.
For $m_\ast=1$ the NGFP is located at
\begin{equation}\label{NGFP}
 \Gt_\ast\approx0.49\;,\quad\Lt_\ast\approx0.17\;,\quad\l_\ast\approx0.44 \; , \quad \Gt_\ast \Lt_\ast \approx 0.085 \, .
\end{equation}
In the compactification limit, $m_\ast\to\infty$ the position of the NGFP asymptotes to
\begin{equation}\label{ngfp1}
 \Gt_\ast\approx0\;,\quad\Lt_\ast\approx0.13\;,\quad\l_\ast\approx0.45 \; .
\end{equation}
For values $m_\ast \le 1$ the position of the NGFP is influenced by the proliferation of singularities in the beta
functions, which will not be investigated further at this stage. Notably, the $m_\ast$ dependence of the NGFP
reveals a turning point at $m_\ast\approx1.35$. Applying the principle of ``minimal sensitivity'', suggests that this point as the most likely value of $m_\ast$.
The NGFP associated with this value is then positioned at
\begin{equation}\label{ngfp2}
 \Gt_\ast\approx0.47\;,\quad\Lt_\ast\approx0.17\;,\quad\l_\ast\approx0.44\; , \quad \Gt_\ast \Lt_\ast \approx 0.078 \, .
\end{equation}

The critical exponents associated with the NGFP are readily computed from the stability matrix \eqref{stab.mat}.
Their $m_\ast$ dependence is shown in the lower panels of Figure \ref{fig:ngfps}. Independently of the actual
value of $m_\ast$ the critical exponents always appear as a real eigenvalue $\theta_1$ and a complex conjugate pair
$\q_\pm=\q'\pm\imath\,\q''$. For the preferred value $m_\ast = 1.35$ these critical exponents  are given by
\begin{equation}
 \q_1\approx2.54\;,\quad\q_\pm \approx 2.30 \pm 1.71\,\imath\; . 
\end{equation}
The corresponding eigenvectors read
\begin{align}
 v_1&\approx \{\, 0.81 \, , \, -0.09 \, , \, 0.58 \,  \} \, , \nonumber\\
 v_\pm&\approx\{ \, 0.88 \, , \, 0.04 \pm 0.45 \imath \, , \, 0.06 \pm 0.01\imath \, \} \, , 
\end{align}
where the numbers refer to the $\Gt$, $\Lt$ and $\l$ direction respectively.
The positive $\theta_1$ and real part of $\theta_\pm$ indicate that the NGFP
is UV attractive in all three eigendirections. As illustrated by Figure 
\ref{fig:ngfps} this property continues to hold for all values $m_\ast \gtrsim 1$.

Interestingly, the universal product $\Gt_\ast \Lt_\ast$ and the critical exponents
of the NGFP are \emph{strikingly similar} to the ones computed within the metric formulation \cite{Reuter:2012id}.
In the present computation, the NGFP is, however, located at $\l_\ast = 0.44$ which is not
compatible with the Einstein-Hilbert action. Nevertheless, we expect that the NGFP
found above is actually the imprint of the NGFP found in the metric formulation
at the level of the foliated flow equation. Its anisotropic position   $\l_\ast \neq 1$
is probably owed to the use of the manifestly anisotropic regulator function $\mathcal{R}_k$ used
in the derivation of the beta function, which gives rise to an extra contribution to the RG flow
perpendicular to the diffeomorphism-invariant hyperplane $\lambda_k = 1$.\footnote{A similar shift 
of the Wilson-Fisher fixed point embedded into the space of anisotropic Lifshitz scalar theories has already been observed in \cite{Bervillier:2004mh}.}  

The fact that the NGFP is UV attractive in all three coupling constants $\Gt$, $\Lt$ and $\l$ makes it tempting
to use this fixed point for the UV completion of the gravitational theory. Provisionally setting $\lambda_* = 1$,
this is the view advocated by the gravitational asymptotic safety scenario. This is, however, \emph{not the viewpoint}
advocated by HL gravity \cite{Horava:2009uw}. The latter case postulates that the UV completion of gravity is provided
by an anisotropic fixed point action containing higher powers of the spatial curvature invariants. These
higher-derivative terms render the theory power-counting renormalizable by producing an anisotropic behavior of the graviton propagator
\begin{equation}
 \frac{1}{\omega^2-\G\left({\bf p}^2\right)^z} \, , 
\end{equation}
$z$ being the dynamical critical exponent $\G$ a coupling constant and $p=(\omega,{\bf p})$ the four-momentum.
The higher-derivative terms that are crucial for encoding this UV behavior are, however, not part of
the ansatz \eqref{ADMaction}. Thus our present computation does not shed light on the proposed UV completion
of HL gravity since the marginal deformations around the 
GFP underlying its perturbative renormalization are not part of the truncation
subspace under consideration. We postpone the analysis to future work, and subsequently focus on the low energy behavior of the theory.

\section{RG flow of anisotropic gravity models II}
\label{sec:flowII}
In Section \ref{finitem} we have found that the ``floating fixed point scenario'', assuming a NGFP value for the Matsubara mass $m$,
does not give rise to the canonical running of the coupling constants expected in the vicinity of the GFP. 
Based on this observation
we advocate the ``interpolating scenario'' where the flow of $m$ actually interpolates between $m_k = m_*$ in the UV and a constant $T =\Tb$ in the IR.
\subsection{Running $m_k$: the interpolating scenario}
The definition of the GFP implies that the dimensionless quantities scale according to their canonical dimension. As a consequence their dimensionful conterparts become scale independent.
 For the Matsubara mass, this behavior should result dynamically from the corresponding beta function $\beta_m$. Since the computation of $\beta_m$ is beyond
the scope of the present paper we construct a model that complements the beta functions for $\Gt$, $\Lt$ and $\lambda$ with a running of the Matsubara mass compatible with
these expectations. Practically, this is achieved by linking the flow of $m_k$ to the running of Newton's constant. While the actual construction may look ad hoc,
it captures the expected running of $m_k$ in the vicinity of the fixed points. Since all properties
uncovered in this section are actually governed by their underlying fixed points, we expect that they are actually independent 
of the precise details of the interpolation. 

The actual construction of $m_k$ proceeds as follows. We start from the dimensionless combination $G_k/T^2$ and assume that
this combination is constant along the RG flow. In terms of the dimensionless quantities, this implies
\begin{equation}
 \p_t\, \left( m^2\Gt \right) = 0\;.
\end{equation}
This relation is easily solved and relates the running of $m_k$ and Newton's constant
\begin{equation}\label{ifpsc}
 m(k) = \frac{2\pi}{\a\sqrt{\Gt(k)}} \, . 
\end{equation}
Here $\a=\Tb/\sqrt{G_0}$ is a free integration constant which measures the length of the ``time'' circle in the IR in Planck units $l_{\rm Pl} \equiv \sqrt{G_0}$.
In the NGFP regime, eq.\ \eqref{ifpsc} fixes $m_\ast$ in terms of $\Gt_\ast$ and the IR parameter $\alpha$
\begin{equation}\label{mvsa}
 m_\ast=\frac{2\pi}{\a\sqrt{\Gt_\ast}} \, . 
\end{equation}
Conceptually, this construction trades the free UV parameter $m_\ast$ for the new parameter $\a$ fixed in the IR region of the theory. The plots in
Figure \ref{fig:ngfps} cover the interval $\a \in [1, 12.1]$. The value $m_\ast = 1.35$ obtained from the principle of minimal sensitivity corresponds to $\alpha = 6.81$ and 
we will resort to this value when constructing numerical solutions in the following subsections.

 Since the ``interpolating scenario'' reduces to the ``floating fixed point scenario'' at high energies,
all results obtained for the NGFP remain valid in this construction. The fact that the scenario also gives rise to the freeze out of the dimensionful coupling constants in the vicinity 
of the GFP can be shown by substituting the relation \eqref{ifpsc} into the system \eqref{msys1} and again performing the linearization
around the GFP. Retaining the terms at leading order in $k$ only this yields 
\be\label{linbetas}
 \b_\Gt\simeq2\Gt \, , \qquad 
 \b_\Lt\simeq-2\Lt + \frac{4}{\pi k\Tb}\,\Gt \, , \qquad 
 \b_\l\simeq-\frac{332}{27\pi k\Tb}\,\Gt\;,
\ee
where besides expanding around the line of GFPs $\Gt_\ast = 0$, $\Lt_\ast =0$ we also expanded around the isotropic fixed point with $\lambda_\ast^{\rm EH} = 1$. 
The numerical factors appearing in the $k$-independent terms coincide with the (inverse) mass dimension of the dimensionful couplings \eqref{dimlessqs},
establishing that $G_k$ and $\Lambda_k$ become scale independent at the GFP. Moreover, the stability coefficients computed from \eqref{linbetas} indicate
that there is one IR attractive, one IR repulsive and one marginal eigendirection.

\subsection{The phase portrait}
\label{sect5.2}
Given the beta functions \eqref{eq:folFinalSystem} supplemented by the running
of the Matsubara mass entailed by the relation \eqref{ifpsc}, we are now in a position
to construct explicit RG trajectories by integrating the \emph{full system} numerically. 
Our primary focus will be on solutions that give rise to a phenomenologically interesting
low energy behavior. As it will turn out, it is the GFP at $\Gt_\ast = 0$, $\Lt_\ast = 0$ 
and $\lambda_\ast = \lambda_\ast^{\rm EH} = 1$ that is crucial for the emergence of a phenomenologically
viable IR behavior of these trajectories.

Before entering into the discussion of the sample trajectories, we make the following crucial observation
concerning the general structure of our phase portrait which can already be concluded from the 
linearized system \eqref{floatlinsys} or  \eqref{linbetas} and also holds for the full beta functions:
the hypersurface $\lambda = 1$ is not invariant under RG flows. This implies that a generic RG trajectory
starting below that plane pierces the $\lambda = 1$ surface at some point and flows to values $\lambda_k > 1$. Typically,
such trajectories run into one of the singular loci depicted in Figure \ref{fig:poles} and terminate at a finite RG scale.
\begin{figure}[!t]
\begin{center}
 \includegraphics[width=10cm,height=8cm]{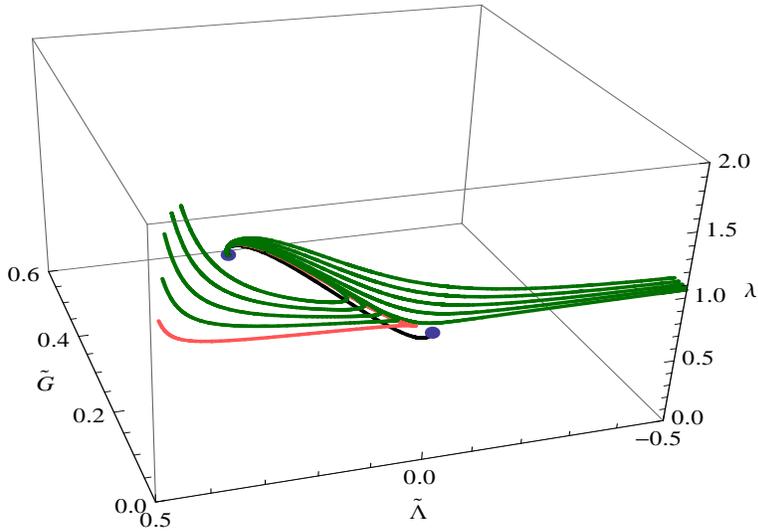}
\end{center}
\caption{A family of well-behaving trajectories that originate from the NGFP (blue point at the left) in the UV and cross over to the GFP (at the right) in the IR.
The two fixed points are connected by a single RG trajectory, the separatrix (black). Depending on whether the other solutions
 flow towards the left or right of the GFP, the theory develops a classical regime with a positive or negative IR value of the cosmological
constant, respectively. All trajectories stay close to the $\lambda = 1$ surface for many orders of RG time $t$. The highlighted trajectory (red) serves as an illustration
for a possible candidate RG trajectory that could realize HL gravity in Nature. Here it was set $\a=6.81$ which corresponds to the value of minimal sensitivity $m_\ast \simeq 1.35$.}
\label{fig:seaweed}
\end{figure}

Instead of classifying all possible RG flows, which would be a laborious task, given the three-dimensional
parameter space of the truncation, we focus on one particular type of RG trajectories, namely 
those which pass close to the GFP with $\lambda_\ast = 1$. A family of such trajectories is shown in 
Figure \ref{fig:seaweed}. These solutions were found by choosing initial conditions close to the GFP and subsequently
tracing their flow by integrating the full system of beta functions numerically towards the UV and IR. 
In the UV, all trajectories are attracted towards the NGFP \eqref{ngfp2}. In the IR we can distinguish three
different behaviors: Firstly, there is a single RG trajectory that connects the NGFP with the GFP. This solution
is called the \emph{separatrix}. It describes a theory that has the classical Einstein-Hilbert action with zero cosmological constant as a low energy limit
\begin{align}\nonumber
 \Gamma_\ast^{\rm EH} &=\frac{1}{16\pi G_0}\int \mathrm d\t \mathrm d^3\!x\,\sqrt{\s}\,\left[K_{ij}K^{ij}-\l_\ast^\textrm{EH}\,K^2-^{(3)\!\!}R\right]\\
 &=\frac{1}{16\pi G_0}\int \mathrm d^4\!x\,\sqrt{g}\,\left(-{}^{(4)}R\right) \, .
\end{align}
Here $G_0$ is a free integration constant which sets the value of Newton's constant in the IR.

Secondly, there are trajectories which start from the NGFP in the UV and cross over to the GFP regime
where they flow towards negative and positive values of $\Lt_k$, respectively. Following the classification
of RG flows in metric gravity \cite{Reuter:2001ag,Rechenberger:2012pm}, we will call these trajectories \emph{generalized Type Ia} and \emph{generalized Type IIIa}.
Their flow is essentially within the hypersurface $\lambda = 1$. As their most important feature,
these trajectories give rise to an extended scaling regime where general relativity with a negative (generalized Type Ia)
and positive (generalized Type IIIa) cosmological constant is a good approximation. This feature becomes increasingly more
pronounced the closer the RG trajectory passes the GFP. The trajectories of generalized Type Ia are well defined for all scales $k \in [0,\infty]$.
Following the flow of the generalized Type IIIa trajectories
towards the deep IR, i.e., beyond the classical regime, the trajectories depart from the $\lambda = 1$ plane and $\lambda_k$ grows rapidly before the trajectories 
terminate in one of the singular loci shown in Figure \ref{fig:poles}. The existence of the separatrix and the RG trajectories of generalized Type Ia and Type IIIa constitutes the main new result of our work. 

The initial conditions chosen to generate the RG trajectories shown in Figure \ref{fig:seaweed} are special in
the sense that the flow passes closely to the GFP with $\lambda_\ast = 1$. From the linearization \eqref{floatlinsys} we concluded
that this fixed point is actually part of the one parameter family of fixed points \eqref{GFP1}. As a consequence plots similar to 
Figure \ref{fig:seaweed} can be obtained for any of the fixed points in this line by choosing initial conditions
so that the flow passes close the the GFP with $\lambda_\ast$. Since the GFP with $\lambda_\ast = 1$ is clearly 
distinguished on phenomenological grounds, we focus on this case in the sequel.

\begin{figure}[!t]
\begin{center}
 \includegraphics[width=5cm,height=3.5cm]{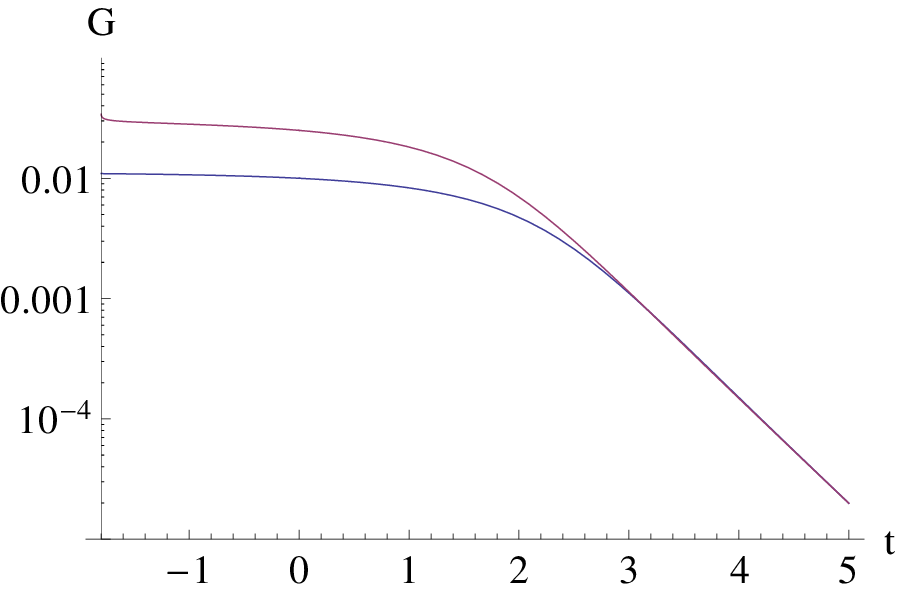}
 \includegraphics[width=5cm,height=3.5cm]{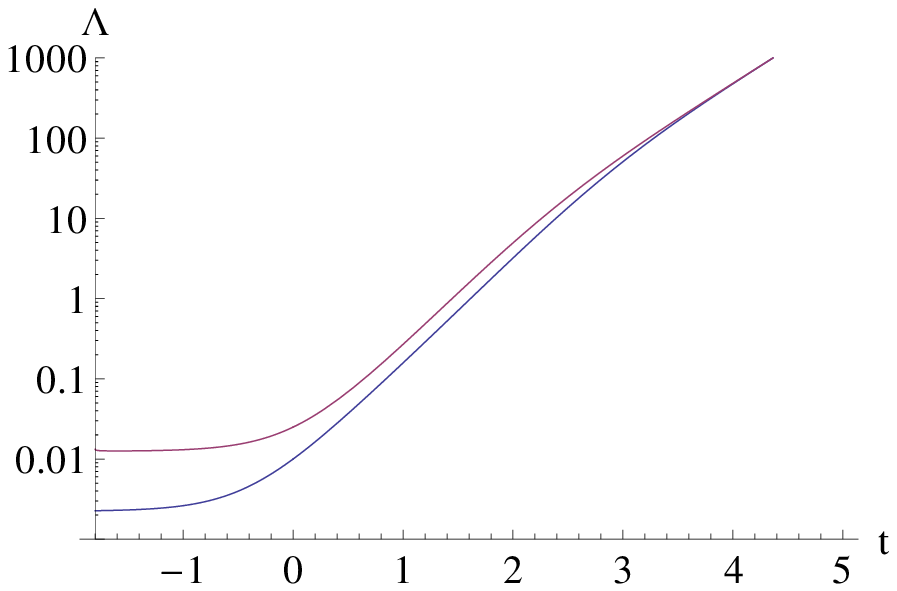}
 \includegraphics[width=5cm,height=3.5cm]{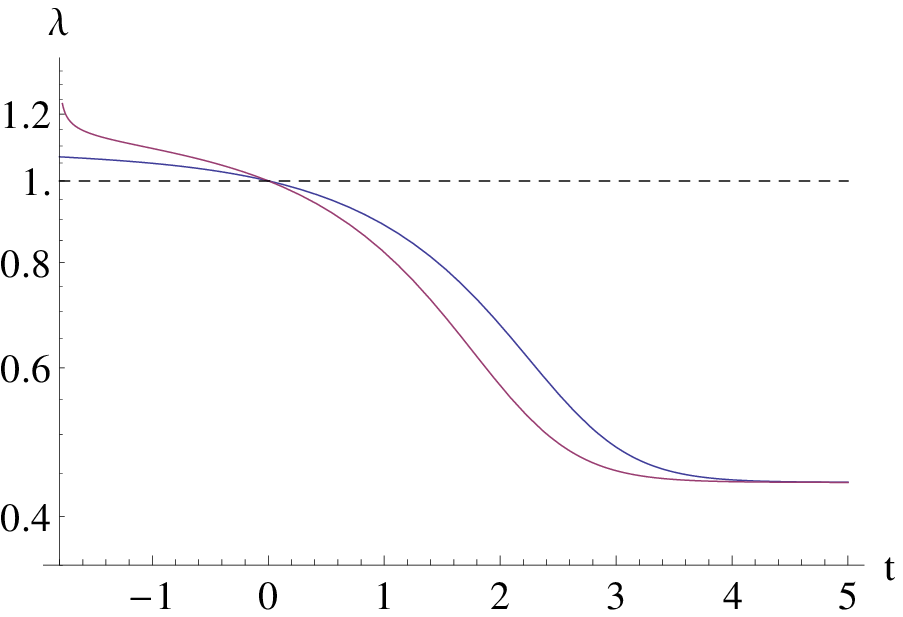}
\end{center}
\caption{RG-time dependence of the dimensionful coupling constants $G_k$ (left), $\Lambda_k$ (middle) and $\l_k$ (right) along typical RG trajectories of generalized Type IIIa. The dashed 
line shown in the third plot indicates the isotropic case. }
\label{fig:tdeps}
\end{figure}

\begin{figure}[!t]
\begin{center}
 \includegraphics[width=5cm,height=3.5cm]{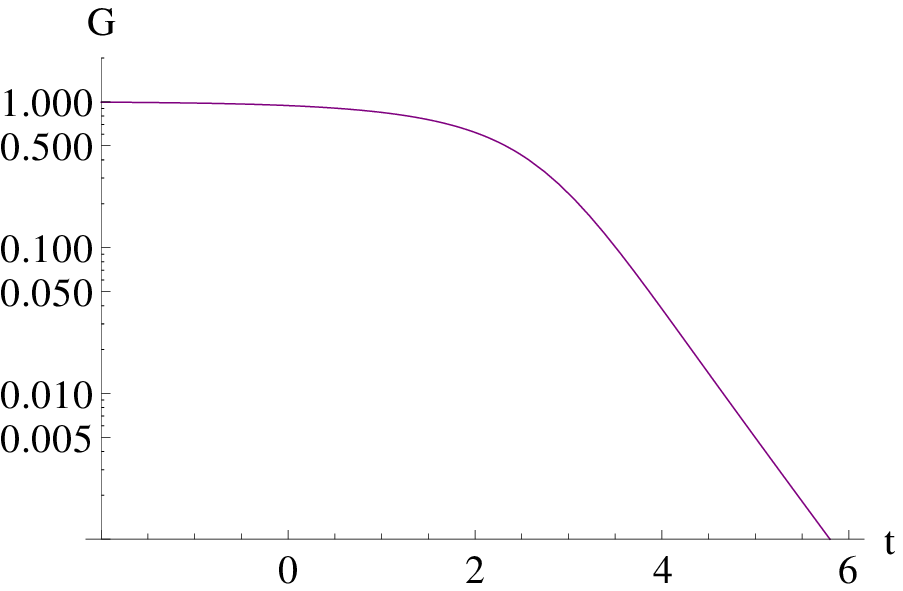}
 \includegraphics[width=5cm,height=3.5cm]{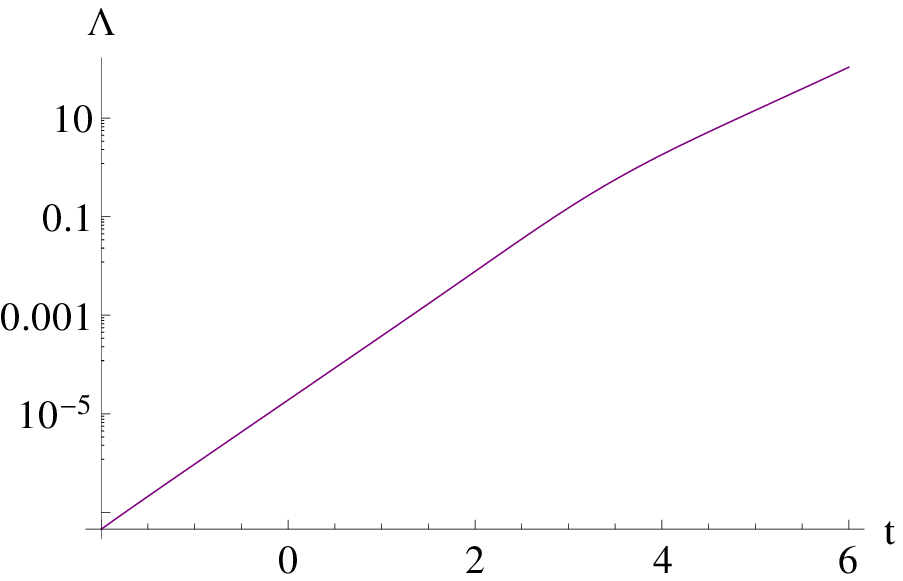}
 \includegraphics[width=5cm,height=3.5cm]{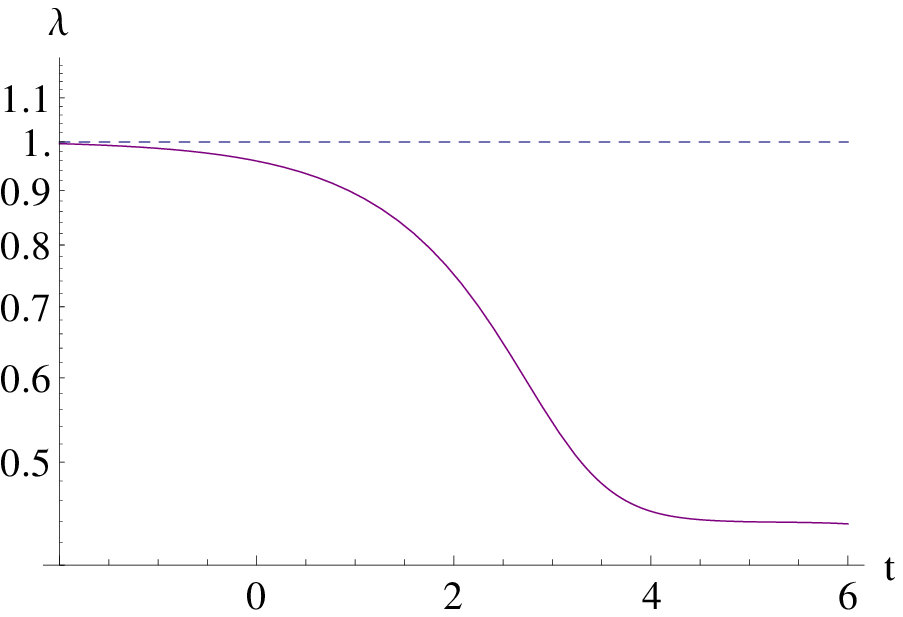}
\end{center}
\caption{RG-time dependence of the dimensionful coupling constants $G_k$ (left), $\Lambda_k$ (middle) and $\l_k$ (right) obtained for the separatrix connecting the NGFP with the GFP with 
$\lambda_\ast^{\rm EH} = 1$. The cosmological constant  vanishes in the low energy limit $\lim_{k \rightarrow 0} \Lambda_k = 0$, while the other couplings approach plateau values
 $G_0 = 1$ and $\lambda_0 = \l_\ast^{\rm EH}$.}
\label{fig:separatrix}
\end{figure}

The existence of the two scaling regimes dominating the RG flow of the
generalized Type IIIa trajectories and the separatrix are 
illustrated further
in Figure \ref{fig:tdeps} and Figure \ref{fig:separatrix}, respectively.
The plots demonstrate that the scale dependence of the couplings is dominated by two scaling regimes
which are connected by a short cross over. For $t > t_{\rm crossover}$ the NGFP induces the scaling 
\be\label{uvcross}
G_k = \Gt_\ast k^{-2} \, , \qquad \Lambda_k = \Lt_\ast k^2 \, , \qquad \lambda_k = \lambda_\ast \, ,
\ee
with the position of the NGFP given by \eqref{ngfp2}. For $t <  t_{\rm crossover}$ the flow is governed by the GFP with $\lambda_\ast = \l_\ast^{\rm EH}$.
The GFP creates plateaus where 
\be\label{scale}
G_k \simeq G_0 \, , \quad \Lambda_k \simeq \Lambda_0 \, , \quad \lambda_k \simeq \lambda_\ast^{\rm EH} \, .
\ee
 The constants $G_0$ and $\Lambda_0$ are determined by the initial conditions specifying the RG trajectory under consideration. 
 For $\Lambda_0 \not = 0$, the classical plateau becomes more pronounced
for smaller IR values $G_0$ (blue curves), i.e., trajectories passing closer to the GFP.
The separatrix shown in Figure \ref{fig:separatrix} thereby constitutes a special case of \eqref{scale} with $\Lambda_0 = 0$.

\subsection{Ho\v{r}ava-Lifshitz gravity realized by Nature}
Given the results of the last subsection, it is interesting to ask
which of the RG trajectories constructed above could be used to
realize HL gravity in Nature. In this subsection we address
this question following the analogous consideration for 
metric gravity \cite{Reuter:2004nx}.

Measurements of the gravitational coupling constants are typically
performed at macroscopic or cosmic distances. For Newton's constant
and the cosmological constant this leads to the observed values
\begin{equation}\label{obsval}
 G_\textrm{obs}=m_\textrm{Pl}^{-2}\, , \qquad \Lambda_\textrm{obs}\approx10^{-122}m_\textrm{Pl}^2\;,
\end{equation}
while astrophysical precision measurements put an upper bound on the deviation of $\lambda$ from
unity \cite{Blas:2009ck,Dutta:2010jh}
\begin{equation}\label{obsl}
 |\l_{\rm obs} -1 |\lesssim4\times 10^{-7}\;.
\end{equation}
The positive value of $\Lambda_{\rm obs}$ thereby indicates that we should consider RG trajectories of generalized Type IIIa. 
The goal is to show that once one adopts \eqref{obsval}, one can find RG trajectories where the scale dependence of $\lambda$ is compatible 
with the bound \eqref{obsl}. 

In terms of RG flows these measurements are carried out in the regime where the trajectories
are well approximated by classical general relativity, i.e., in the realm of the GFP with $\l_\ast = \l_\ast^{\rm EH} = 1$.
In this region the flow has a good description in terms of the linearized system \eqref{linbetas}.
This system is readily solved analytically, yielding 
\be\label{linsoldimless}
 \Gt(k) = C_1\,k^2 \, , \qquad  \Lt(k) = \frac{C_2}{k^2}+\frac{4}{3\pi}\,\frac{k\,C_1}{\bar{T}} \, , \qquad \l(k) = C_3-\frac{332}{27\pi}\,\frac{k\,C_1}{\bar{T}} \, . 
\ee
Here the $C_i$ are integration constants that specify the RG trajectory under consideration. In order 
to relate these constants to the observed quantities, we express this solution in terms of the dimensionful
couplings
\be
 G(k) =C_1 \, , \qquad 
 \Lambda(k) =C_2+\frac{4}{3\pi}\,\frac{k^3C_1}{\bar{T}} \, , \qquad 
 \l(k) =C_3-\frac{332}{27\pi}\,\frac{k\,C_1}{\bar{T}}\;.
\ee
Imposing that the measurements \eqref{obsval} are essentially carried out at $k = 0$ fixes the constants $C_1 = G_{\rm obs}$ and $C_2 = \Lambda_{\rm obs}$,
which we complement by $C_3 = 1$.\footnote{Here we made the implicit assumption that the ``classical regime'' associated with the GFP actually extends
all the way down to $k=0$, thereby neglecting the possibility IR modifications of general relativity. This assumption is justified by the observation that for a wide range of scales (approximately 20 orders of magnitude) one has $\Lambda_\textrm{obs}\gg k^3G_\textrm{obs}/\Tb$ and as a consequence $\Lambda(k)\simeq\Lambda_\textrm{obs}$, yet $\Lt(k)\ll1$ so that the departure from the linearized behavior is still negligible.}

Our goal is to estimate the running of $\l(k)$ in the classical regime between $k=0$ and a scale $k_{\rm tp}$. Based on the flow pattern 
in Figure \ref{fig:seaweed} we follow the idea of \cite{Reuter:2004nx} and take $k_{\rm tp}$ to be the turning point appearing in the flow of $\Lt$ satisfying
$\p_k \Lt_k|_{k = k_{\rm tp}} = 0$. Based on the solution \eqref{linsoldimless}, it is straightforward to determine
\begin{equation}
 k_\textrm{tp}=\sqrt[3]{\frac{3\pi\,\bar{T}\Lambda_\textrm{obs}}{2\,G_\textrm{obs}}}
\end{equation}
For $\alpha = 6.81$ the turning point is situated at $k_{\rm tp} = 6.84 \times 10^{-41} \, m_{\rm Pl}$, which corresponds to typical terrestrial distances.
Evaluating the relations \eqref{linsoldimless} at this scale gives the corresponding value of the dimensionless coupling constants
\begin{equation}\label{tppos}
 \Gt(k_\textrm{tp})\approx4.69\times10^{-81}\quad,\qquad\Lt(k_\textrm{tp})\approx6.40\times10^{-42}\quad,\qquad\l(k_\textrm{tp})-1\approx-3.93\times10^{-41}\;.
\end{equation}
The important result here is the actual minuscule running of $\l$ in the GFP regime: The scale dependence implied by \eqref{tppos} is well below the observational bound \eqref{obsl}. 

The mechanism underlying this ``quenching'' of the running of $\l$ is rather generic and simple. The RG trajectory corresponding to the observational values
\eqref{obsval} is not generic: as illustrated by the values \eqref{tppos} it is highly fine-tuned in the sense that it passes incredibly close to the GFP. This property
is required for obtaining the observed value of the cosmological constant and also leads to a dramatic suppression of the running of $\lambda$. This leaves the GFP
as a viable mechanism for constructing RG flows in HL gravity which are compatible with observations.

\section{Conclusions}
\label{sec:con}
In this work, we have used the functional renormalization group equation (FRGE) for projectable Ho\v{r}ava-Lifshitz gravity \cite{Manrique:2011jc,Rechenberger:2012dt} to study the theories RG flow at low energies. This FRGE constitutes the natural generalization of the FRGE for metric gravity \cite{mr} to the case of foliated spacetimes. The symmetry group implemented in its construction consists of foliation-preserving diffeomorphisms Diff$_{\cF}(\cM, \Sigma)$ which form a subgroup of the full diffeomorphism group underlying general relativity and the metric formulation of the FRGE. As a consequence, the RG flow of HL gravity contains additional coupling constants that in the metric formulation are fixed by the additional symmetries. Thus the theory space of metric gravity forms a natural subspace of the one underlying HL gravity, see \cite{Rechenberger:2012dt} for a detailed discussion.

Our ansatz \eqref{ADMaction} for the effective average action $\Gamma_k$ includes a scale-depended Newton's constant $G$, the cosmological constant $\Lambda$ and anisotropy parameter $\lambda$. The beta functions for these running couplings are computed explicitly by substituting this ansatz into the FRGE. Based on these beta functions, we identified
 the line of Gaussian fixed points \eqref{GFP1}. This line includes the Einstein-Hilbert action with vanishing cosmological constant for $\lambda_* = 1$. 
This fixed point constitutes a saddle point for the RG flow with one IR repulsive, one IR attractive and one marginal eigendirection. Moreover, the beta functions possess a nontrivial fixed point, which is UV attractive for all three coupling constants included in our ansatz. Based on its critical exponents, we expect that this fixed point is actually the imprint of the (isotropic) nontrivial UV fixed point underlying the gravitational Asymptotic Safety program \cite{Niedermaier:2006wt,Reuter:2007rv,Percacci:2007sz,Litim:2008tt,Reuter:2012id}. The numerical construction of RG trajectories displayed in Figure \ref{fig:seaweed} furthermore shows that there is a crossover between the UV fixed point and the GFP in the IR where general relativity constitutes a good approximation. This flow pattern is actually very reminiscent of the one found in the $R^2$ truncation of metric gravity \cite{Rechenberger:2012pm}, where the coupling constants have the same dimensionality as in the present case.

The RG flow constructed in this paper actually shows two \emph{a priori} unexpected features. Firstly, the isotropic plane with $\lambda = 1$ is not a fixed plane of the RG flow, \emph{i.e.} a RG flow starting on the symmetric plane is driven towards larger values of the anisotropy parameter $\lambda > 1$ when traced towards the IR.
Similarly, the plane $\lambda = 1/3$, which is distinguished by a classical decoupling of the fluctuations in the conformal mode, also does not 
give rise to special properties with respect to the RG flow. Secondly, the non-Gaussian fixed point actually located at $\lambda_* = 0.44$ thus does not seem to correspond to an isotropic field theory. We expect that this features is related to the particular choice of the cutoff $\R_k$ entering into the construction of the FRGE. Motivated by the analogy to quantum field theory at finite temperature we chose a regulator that is purely spatial. Thus the present formulation of the FRGE contains two pieces that actually source violations of the 
diffeomorphism invariance: anisotropic couplings in the gravitational part of $\Gamma_k$ and the regulator. We expect that 
it is actually the latter that is responsible for the shift of the NGFP away from $\lambda_\ast = 1$.
In the context of anisotropic scalar field theories, a similar observation has already been made in \cite{Bervillier:2004mh}, where the Wilson-Fisher fixed point was converted to 
an anisotropic fixed point, probably due to the use of an anisotropic regulator. Let us stress, however, that we do not expect that the GFP is affected by these effects, since this point appears as an endpoint of an RG trajectory ($k\to0$) where the regulators disappear. Clarifying the position of the nontrivial fixed point and the RG properties of the $\lambda = 1$ hyperplane will, presumably, require the rederivation of the present beta functions based on a symmetric regulator. This is, however, beyond the scope of the present work, and we hope to come back to this point in the future. 

At this stage, we also stress that the free-field fixed points, supposed to provide the UV completion of HL gravity \cite{Horava:2009uw}, cannot be probed by the low energy ansatz \eqref{ADMaction}. The study of the UV properties of the theory will require supplementing this truncation by the remaining relevant and marginal operators of the theory and study the flow of
\be
\Gamma_k^{\rm grav} = \int \mathrm d\t \mathrm d^3x N \sqrt{\sigma} \sum_{[\cO_i] = 6} g_{i,k} \cO_i + 
\int \mathrm d\t \mathrm d^3x N \sqrt{\sigma} \sum_{[\cO_j] < 6} g_{j,k} \cO_j \, . 
\ee
In the case that the free-field UV fixed point indeed satisfies the detailed balance condition (which is not strictly necessary) a suitable ansatz for probing the renormalizability of the theory beyond power-counting arguments approximates $\Gamma_k$ by
\be
\Gamma_k = \int \mathrm d\t \mathrm d^3x N \sqrt{\sigma} \left[ \frac{1}{16 \pi G_k} \left( K_{ij} K^{ij} - \lambda_k K^2\right) - \frac{16 \pi G_k}{w_k^4} C_{ij} C^{ij} \right]
\ee
with $C_{ij}$ being the three-dimensional Cotton tensor. The condition that the interactions are switched off corresponds to sending $w \rightarrow 0$, so that the free-field fixed points are located on the two-dimensional subspace spanned by the marginal couplings $\lambda$ and $\gamma \equiv G/w$.
Based on the example of higher derivative gravity \cite{Codello:2006in,Niedermaier:2009zz,Groh:2011vn} one expects that the beta functions for $\lambda$ and $\gamma$ single out discrete points on this hypersurface which, once the flow indeed turns out to be asymptotically free, could constitute suitable UV fixed points of gravity. In this picture the interaction monomials studied in this paper constitute relevant deformations of such a free-field fixed point. Their inclusion triggers the flow away from the free-field fixed point in the UV to the IR of the theory. For suitable initial conditions the latter could be provided by the GFP discovered in this paper where general relativity constitutes a good approximation.

The phase portrait shown in Figure \ref{fig:seaweed} suggests that the IR behavior of both HL gravity and Asymptotic Safety is controlled by the same GFP. In this regime, the two theories lead to very distinct signatures: in HL gravity, the RG flow towards a positive cosmological constant are accompanied by a tiny (but nonzero) deviation of the anisotropy parameter $\lambda$ away from its value in general relativity. In Asymptotic Safety on the other hand the anisotropy parameter $\lambda = 1$ is protected by the symmetries of the theory so that no such effect will be present. On this basis, we expect that it is possible to distinguish between the two candidate UV completions of gravity on an experimental basis.

In summary, we expect that the FRGE \cite{Manrique:2011jc,Rechenberger:2012dt} provides a valuable analytic complement to quantizing HL gravity by discrete Monte Carlo methods \cite{Ambjorn:2010hu,Anderson:2011bj,Ambjorn:2013tki} along the lines of the Causal Dynamical Triangulations (CDT) program \cite{Ambjorn:2012jv}. We hope to come back to this point in the future.

\section*{Acknowledgements}

The work of A.C., S.R., and F.S.\ is supported by the Deutsche Forschungsgemeinschaft (DFG)
within the Emmy Noether program (Grant SA/1975 1-1). S.R. also acknowledges support by HIC for FAIR within the LOEWE program of the State of Hesse during the final stages of this work.

\begin{appendix}
\section{The scalar trace}
 \label{app:scalarTrace}
In this appendix we evaluate the scalar trace \eqref{eq:foliatedS}. It reads
\begin{equation}
\mathcal T_\mathrm{scalar} = \frac{1}{2} \mathrm{Tr} \Big[ \frac{\bar\Gamma_{\varsigma\varsigma}^{(2)}\partial_t\mathcal R_k^{hh} + \bar\Gamma_{hh}^{(2)}\partial_t\mathcal R_k^{\varsigma\varsigma} - 2 \bar\Gamma_{h\varsigma}^{(2)}\partial_t\mathcal R_k^{h\varsigma}}{\bar\Gamma_{hh}^{(2)} \bar\Gamma_{\varsigma\varsigma}^{(2)} - \bar\Gamma_{h\varsigma}^{(2)} \bar\Gamma_{\varsigma h}^{(2)}} \Big]
\end{equation}
with the second variations given in \eqref{2ndvariations3} and the regulators given in \eqref{eq:foliatedRegulator}.

Let us also state that we are not interested in the complete trace, but only in constant terms, terms proportional to the background scalar curvature $\bar R$ and terms proportional to $\gamma^2$. Thus we use an expansion for $\bar R$ and $\gamma$ and omit all higher-order terms. Therefore we start to investigate the denominator which is a function of both parameters
\begin{equation}
\frac{1}{16\pi G_k} f(\bar R, \gamma) = \bar\Gamma_{hh}^{(2)} \bar\Gamma_{\varsigma\varsigma}^{(2)} - \bar\Gamma_{h\varsigma}^{(2)} \bar\Gamma_{\varsigma h}^{(2)} \; .
\end{equation}
In order to expand the expression under the trace we have to calculate $f(0,0), \partial_{\bar R} f(0,0), \partial_\gamma f(0,0)$ and $\partial_\gamma^2 f(0,0)$. These can be evaluated to
\begin{align}
f(0,0) = f_{00}^1 P_k + f_{00}^2 \, , \qquad
\partial_{\bar R} f(0,0) = f_{10}^1 P_k + f_{10}^2 \, , \nonumber \\
\partial_\gamma f(0,0) = f_{01}^1 P_k + f_{01}^2 \, , \qquad
\partial_\gamma^2 f(0,0) = f_{02}^1 P_k + f_{02}^2
\end{align}
with the prefactors $f_{pq}^r$ given in terms of the dimensionless quantities
\begin{equation}
\Lt_k = \frac{\Lambda_k}{k^2} \, , \qquad m = \frac{2\pi}{Tk} \, , \qquad \Gt_k = G_k k^2 \, , \qquad \lambda_k \, .
\end{equation}
Explicitly the prefactors read
\begin{align}
f_{00}^1 =& \frac{2k^2}{9}\left(\Lt_k - (1-\lambda_k)m^2n^2 \right) \, , \nonumber \\
f_{00}^2 =& \frac{2k^4}{9}\left(\Lt_k - (1-\lambda_k)m^2n^2\right) w_0 \, , \nonumber \\
f_{10}^1 =& -\frac{1}{27}\, , \nonumber \\
f_{10}^2 =& k^2\frac{1-2\lambda_k}{9}m^2n^2 \, , \nonumber  \\
f_{01}^1 =& \imath k \frac{5-9\lambda_k}{9} mn \, , \nonumber \\
f_{01}^2 =& -\imath k^3 \left(\frac{4(3\l_k-1)(3\l_k-2)}{9}m^3n^3+\frac{5-6\l_k}{9}\Lt_kmn\right) \, , \nonumber \\
f_{02}^1 =& -\frac{19(3\l_k-1)}{18} \, , \nonumber \\
f_{02}^2 =& k^2\left(\frac{2-6\l_k}{9}\Lt_k-\frac{(1-3\l_k)(-90\l_k+49)}{18}m^2n^2\right)
\end{align}
with
\begin{equation}\label{eq:w0}
 w_0=-\frac{2\Lt_k^2+(1-6\l_k)\Lt_km^2n^2-(1-3\l_k)m^4n^4}{\Lt_k-(1-\l_k)m^2n^2} \, .
\end{equation}
This can be used to perform the expansion in $\bar R$ and $\gamma$
\begin{align}
\frac{1}{a+b\bar R} &= \frac{1}{a} - \frac{b\bar R}{a^2} + O(\bar R^2) \, , \nonumber \\
\frac{1}{a + b\gamma + c\gamma^2} &= \frac{1}{\gamma} - \frac{b\gamma}{a^2} + \frac{b^2-ac}{a^3}\gamma^2 + O(\gamma^3) \; .
\end{align}
Furthermore we need
\begin{equation}
\frac{1}{(a+b\gamma+c\gamma^2)^2} = \frac{1}{a^2} + O(\gamma) \; .
\end{equation}
To evaluate these expansions systematically we decompose $\mathcal T_\mathrm{scalar}$, for convenience, according to
\begin{equation}
\mathcal T_\mathrm{scalar} = \mathcal T_\mathrm{scalar}^\mathrm{I} + \mathcal T_\mathrm{scalar}^\mathrm{II} + \mathcal T_\mathrm{scalar}^\mathrm{III} \; .
\end{equation}
These three parts can be expanded separately in $\bar R$ and $\gamma$ to give
\begin{align} \label{eq:foliatedScalarExpansion}
\mathcal T_\mathrm{scalar}^\mathrm{I} &= \frac{1}{2} \mathrm{Tr} \Big[ \frac{\bar\Gamma_{\varsigma\varsigma}^{(2)}\partial_t\mathcal R_k^{hh}}{\bar\Gamma_{hh}^{(2)} \bar\Gamma_{\varsigma\varsigma}^{(2)} - \bar\Gamma_{h\varsigma}^{(2)} \bar\Gamma_{\varsigma h}^{(2)}} \Big] \nonumber \\
&\!\!\!= -\frac{1}{9} \sum_n \mathrm{tr} \Big\{ \tfrac{\partial_t(T Z_{\mathrm Nk} R_k)}{T Z_{\mathrm Nk}}\Big[ \tfrac{-\tfrac{2}{9} P_k + b_1}{f(0,0)} + \bar R\Big( \tfrac{b_2}{f(0,0)} - \tfrac{-\tfrac{2}{9} P_k + b_1}{f(0,0)^2}\partial_{\bar R}f(0,0) \Big) \nonumber \\
& \!\!\!+\gamma^2 \Big( \tfrac{2b_4}{f(0,0)} - \tfrac{b_3}{f(0,0)^2}\partial_\gamma f(0,0) + \tfrac{-\tfrac{2}{9} P_k + b_1}{2}\Big( -\tfrac{\partial_\gamma^2 f(0,0)}{f(0,0)^2} + 2\tfrac{(\partial_\gamma f(0,0))^2}{f(0,0)^3} \Big) \Big) \Big] \Big\} \; , \nonumber \\
\mathcal T_\mathrm{scalar}^\mathrm{II} &= \frac{1}{2} \mathrm{Tr} \Big[ \frac{\bar\Gamma_{hh}^{(2)}\partial_t\mathcal R_k^{\varsigma\varsigma}}{\bar\Gamma_{hh}^{(2)} \bar\Gamma_{\varsigma\varsigma}^{(2)} - \bar\Gamma_{h\varsigma}^{(2)} \bar\Gamma_{\varsigma h}^{(2)}} \Big] \nonumber \\
&\!\!\!=-\frac{1}{9} \sum_n \mathrm{tr} \Big\{ \tfrac{\partial_t(T Z_{\mathrm Nk} R_k)}{T Z_{\mathrm Nk}}\Big[ \tfrac{-\tfrac{2}{9} P_k + b_5}{f(0,0)} + \bar R\Big( \tfrac{b_6}{f(0,0)} - \tfrac{-\tfrac{2}{9} P_k + b_5}{f(0,0)^2}\partial_{\bar R}f(0,0) \Big) \nonumber \\
& \!\!\!+\gamma^2 \Big( \tfrac{2b_8}{f(0,0)} - \tfrac{b_7}{f(0,0)^2}\partial_\gamma f(0,0) + \tfrac{-\tfrac{2}{9} P_k + b_5}{2}\Big( -\tfrac{\partial_\gamma^2 f(0,0)}{f(0,0)^2} + 2\tfrac{(\partial_\gamma f(0,0))^2}{f(0,0)^3} \Big) \Big) \Big] \Big\} \; , \nonumber \\
\mathcal T_\mathrm{scalar}^\mathrm{III} &= \frac{1}{2} \mathrm{Tr} \Big[ \frac{ -2 \bar\Gamma_{h\varsigma}^{(2)}\partial_t\mathcal R_k^{h\varsigma}}{\bar\Gamma_{hh}^{(2)} \bar\Gamma_{\varsigma\varsigma}^{(2)} - \bar\Gamma_{h\varsigma}^{(2)} \bar\Gamma_{\varsigma h}^{(2)}} \Big] \nonumber \\
&= - \tfrac{4}{81}\sum_n \mathrm{tr} \Big\{ \tfrac{\partial_t(T Z_{\mathrm Nk} R_k)}{T Z_{\mathrm Nk}}\Big[ \tfrac{P_k}{f(0,0)} - \bar R \Big( \tfrac{1}{4f(0,0)} + \tfrac{P_k \partial_{\bar R}f(0,0)}{f(0,0)^2} \Big) \nonumber \\
&+ P_k\gamma^2 \Big( -\tfrac{\partial_\gamma^2f(0,0)}{f(0,0)^2} + 2\tfrac{(\partial_\gamma f(0,0))^2}{f(0,0)^3} \Big) \Big] \Big\}
\end{align}
where we omitted irrelevant terms and introduced for better readability the abbreviations
\begin{align}
b_1 &= -\frac{4}{3}\Lt_k k^2 + \frac{2}{3}m^2n^2k^2 \, ,&
b_5 &= \frac{1}{3}\Lt_k k^2 + \frac{1-3\lambda_k}{3}m^2n^2k^2 \, , \nonumber \\
b_2 &= \tfrac{2}{9} \, ,&
b_6 &= \frac{1}{18} \, , \nonumber \\
b_3 &= \imath\frac{12\lambda_k - 7}{3}mnk \, ,&
b_7 &= \imath \frac{3\lambda_k-1}{6}mnk \, , \nonumber \\
b_4 &= \frac{11(3\lambda_k -1)}{6} \, ,&
b_8 &= \frac{13(3\lambda_k-1)}{24} \, ,
\end{align}
that should not be confused with the abbreviations introduced in the main text. We used the same symbols here, since a confusion is very unlikely.

Now all traces appearing in \eqref{eq:foliatedScalarExpansion} are of the form discussed in Appendix A of \cite{Rechenberger:2012dt}. Thus we can evaluate the traces and combine the result for $\mathcal T_\mathrm{scalar}^\mathrm{I}, \mathcal T_\mathrm{scalar}^\mathrm{II}$ and $\mathcal T_\mathrm{scalar}^\mathrm{III}$. The result is very lengthy and thus again we introduce abbreviations, labeled with $a_k, k=1, \ldots, 14$, to improve the readability. The result, after evaluating the trace, reads
\begin{align} \label{eq:TscalarApp}
\mathcal T&_\mathrm{scalar} = -\frac{2k^3}{(4\pi)^{3/2}} \sum_n \int \mathrm d^3\!x\,\sqrt{\sb}\, \bigg[ a_1 q_{3/2}^{1,0} + \frac{\bar R}{k^2} \Big( \tfrac{a_1}{6}q_{1/2}^{1,0} + a_2Xq_{3/2}^{1,0} + (a_3+a_4m^2n^2)Xq_{3/2}^{2,0} + a_5Xq_{3/2}^{2,-1} \Big) \nonumber \\
&+ \frac{\gamma^2}{k^2} \Big( a_6Xq_{3/2}^{1,0} + (a_7+a_8m^2n^2+a_9Xm^2n^2+a_{10}Xm^4n^4)Xq_{3/2}^{2,0}  + (a_{11}+a_{12}Xm^2n^2)Xq_{3/2}^{2,-1} \nonumber \\
& + (a_{13}+a_{14}m^2n^2+a_{15}m^4n^4)X^2m^2n^2q_{3/2}^{3,0}+ (a_{16}+a_{17}m^2n^2)X^2m^2n^2q_{3/2}^{3,-1} + a_{18}X^2m^2n^2q_{3/2}^{3,-2} \Big) \bigg] \; . 
\end{align}
Here all $q$ functions are understood to be evaluated at $w_0$, given in \eqref{eq:w0}, 
and the prefactors read
\begin{align} \label{eq:foliatedScalarPrefactors}
X &= \left( \Lt_k - (1-\lambda_k)m^2n^2 \right)^{-1} \, , \nonumber \\
a_1 &= -\tfrac{1}{2} \, , \quad a_2 = \tfrac{1}{12} \, , \quad a_3 = 0 \, , \quad a_4 = \tfrac{1-2\lambda_k}{4} \, , \quad a_5 = -\tfrac{1}{12} \, , \nonumber \\
a_6 &= \tfrac{19(3\lambda_k-1)}{16} \, , \qquad a_7 = \tfrac{(1-3\l_k)\Lt_k}{4} \, , \nonumber \\
a_8 &= \tfrac{(3\lambda_k-1)(49-90\l_k)}{16} \, , \nonumber \\
a_9 &= \tfrac{(5-9\l_k)(5-6\l_k)\Lt_k}{8} \, , \nonumber \\
a_{10} &= -\tfrac{(3\l_k-1)(3\l_k-2)(9\l_k-5)}{2} \, , \qquad a_{11} = -\tfrac{19(3\l_k-1)}{16} \, , \nonumber \\
a_{12} &= -\tfrac{\left(5-9\lambda_k\right)^2}{8} \, , \qquad a_{13} = \tfrac{(5-6\l_k)^2\Lt_k^2}{8} \, , \nonumber \\
a_{14} &= (5-6\l_k)(3\l_k-1)(3\l_k-2)\Lt_k \, , \qquad a_{15} = 2(3\l_k-1)^2(3\l_k-2)^2 \, , \nonumber \\
a_{16} &= -\tfrac{(5-6\l_k)(5-9\l_k)\Lt_k}{4} \, , \quad a_{17} = (5-9\l_k)(1-3\l_k)(3\l_k-2) \, , \nonumber \\
a_{18} &= \tfrac{(5-9\l_k)^2}{8} \; .
\end{align}
The scalar trace \eqref{eq:TscalarApp} constitutes the final result of this appendix. Added to the transverse traceless trace (\ref{expandedTT}) it gives the right hand side of the foliated Wetterich equation \eqref{wettericheqn}.

\section{Summing the Matsubara Modes}\label{app:MatsubaraSum}

This appendix treats the sums over Matsubara modes which are left after performing the operator traces. Due to simplicity we start with the transverse traceless sector, before turning to the Matsubara sums in the scalar sector. The sums appearing in \eqref{eq:TTT} are of the form $\sum_n (m^2n^2)^r q^{p,q}_l(w_\mathrm{TT})$ where
\begin{equation} \label{eq:w2Tapp}
w_\mathrm{TT} = -2\Lt_k +m^2n^2 \; .
\end{equation}
The $q$ functions have been defined in Section \ref{sec:betas} and read
\begin{equation}
q^{p,q}_l(w) = \Phi^{p,q}_l(w) - \tfrac{1}{2} \eta_N \tilde{\Phi}^{p,q}_l(w) \; , \qquad \eta_\mathrm N = -\partial_t \ln(T\,Z_{\mathrm Nk}) \; .
\end{equation}
The threshold functions $\Phi^{p,q}_l$ are given in Appendix A of \cite{Rechenberger:2012dt} and we specify to the simple result \eqref{eq:optimisedThreshold} reached for the optimised cutoff. Therefore all sums are of the form $\sum_n n^{2r}(1+w_\mathrm{TT})^{-p}$ with $w_\mathrm{TT}$ given in \eqref{eq:w2Tapp}. The infinite sums can be performed by utilising the master formulas
\begin{equation} \label{eq:mastersum}
\sum_{n = - \infty}^\infty \frac{1}{n^2 + x^2} =  \frac{\pi}{x \, \tanh(\pi x)} \, , \qquad
\sum_{n = - \infty}^\infty \frac{1}{n^2 - x^2} =  - \frac{\pi}{x \, \tan(\pi x)} \, .
\end{equation}
To shorten the notation we introduce the following summed threshold functions for $r\leq p-1$
\begin{equation}
\Upsilon_{l}^{p,r} = \!\! \sum_{n = - \infty}^\infty \!\! \left( m^2 n^2 \right)^r \, \Phi_{l}^{p,q}(w_\mathrm{TT}) \; , \qquad
\tilde{\Upsilon}_{l}^{p,r} = \!\! \sum_{n = - \infty}^\infty \!\! \left( m^2 n^2 \right)^r \, \tilde{\Phi}_{l}^{p,q}(w_\mathrm{TT}) \; .
\end{equation}

There are only few combinations of $p$ and $r$ appearing in \eqref{eq:TTT}, which are $(p,r) = (1,0), (2,0), (3,1)$. The summation of $\Upsilon_{l}^{1,0}$ is fairly simple since it is a direct implementation of the master formula \eqref{eq:mastersum}. The evaluation gives
\begin{equation}
\Upsilon_{l}^{1,0} = \frac{1}{\Gamma(l+1)} \frac{\pi}{m} \, \sqrt{\frac{1}{1-2\Lt_k}} \, \frac{1}{\tanh\left( \tfrac{\pi}{m} \sqrt{1-2\Lt_k}  \right) } \; .
\end{equation}
The next summed threshold functions, $\Upsilon_{l}^{p,0}, p>1$, can be deduced iteratively from the first one by considering $\Upsilon_{l}^{p+1,0} = \frac{1}{2p}\partial_{\Lt_k}\Upsilon_{l}^{p,0}$. This leads to
\begin{equation}
\Upsilon_{l}^{2,0} =  \, \tfrac{1}{\Gamma(l+1)} \tfrac{\pi}{2m(1-2\Lt_k)} \left( 
\tfrac{\left( \sqrt{1-2\Lt_k} \, \right)^{-1}}{\tanh\left( \tfrac{\pi}{m} \sqrt{1-2\Lt_k}  \right)} +  \tfrac{\pi}{m \, \sinh\left( \tfrac{\pi}{m} \sqrt{1-2\Lt_k}  \right)^2}
\right) \; .
\end{equation}
Terms with higher values for $r$ can be found recursively via $\Upsilon_{l}^{p+1,r+1} = -\frac{1}{2pm}(m^2)^{r+1}\partial_m(\Upsilon_{l}^{p,r}/(m^2)^r)$. This iteration gives us the last missing summed threshold function, which reads
\begin{align}
\Upsilon_{l}^{3,1} = \tfrac{1}{\Gamma(l+1)}\tfrac{\pi}{16m^3}\left( \tfrac{1}{1-2\Lt_k} \right)^{\frac{3}{2}} \!\! \bigg( &\tfrac{2m^2}{\tanh\left( \tfrac{\pi}{m} \sqrt{1-2\Lt_k}  \right)} + \tfrac{2\pi m\sqrt{1-2\Lt_k}}{\sinh\left( \tfrac{\pi}{m} \sqrt{1-2\Lt_k}  \right)^2}- \tfrac{4\pi^2(1-2\Lt_k)}{\tanh\left( \tfrac{\pi}{m} \sqrt{1-2\Lt_k}  \right) \sinh\left( \tfrac{\pi}{m} \sqrt{1-2\Lt_k}  \right)^2} \bigg) \; .
\end{align}
The threshold functions denoted with a tilde are related to the ones above by $\tilde\Upsilon_{l}^{p,r} = \frac{\Gamma(l+1)}{\Gamma(l+2)}\Upsilon_{l}^{p,r}$. Finally, for convenience, we introduce the summed version of the $q$ functions which we denote by
\begin{equation} \label{eq:summedq2T}
T^{p,r}_l = {\Upsilon}^{p,r}_l - \tfrac{1}{2} \, \eta_N \, \tilde{\Upsilon}^{p,r}_l \; .
\end{equation}

Similar considerations have to be done for the scalar part. However this is more involved since $w_0$ is a polynomial of second order in $n^2$ and thus the master formula \eqref{eq:mastersum} can not be applied directly. To boil the appearing terms down to the form \eqref{eq:mastersum} we use the partial fraction decomposition. Therefore we write
\begin{align}
1+w_0 &= \left. \frac{\kappa_1\Lt_k -2\kappa_2\Lt_k^2 - \left( \kappa_3(1-6\l_k) \Lt_k + \kappa_1(1 - \lambda_k) \right) m^2n^2 + \kappa_4(1-3\lambda_k)m^4n^4}{\kappa_5\Lt_k - (1 - \lambda_k)m^2n^2} \right|_{\kappa_1=\ldots=\kappa_5=1} \nonumber \\
&= \left. \frac{\kappa_4(1-3\lambda_k)m^4(n^2-w_+)(n^2-w_-)}{\kappa_5\Lt_k - (1 - \lambda_k)m^2n^2} \right|_{\kappa_1=\ldots=\kappa_5=1}
\end{align}
where we introduced five auxiliary parameters $\kappa_1,\ldots,\kappa_5$. The zeros $w_\pm$ are given as
\begin{equation} \label{eq:wpm}
w_\pm = \frac{\kappa_3(1-6\l_k)\Lt_k + \kappa_1(1-\lambda_k)}{2\kappa_4(1-3\lambda_k)m^2}\pm \sqrt{ \frac{\left(\kappa_3(1-6\l_k)\Lt_k + \kappa_1(1-\lambda_k) \right)^2}{4\kappa_4^2(1-3\lambda_k)^2m^4} - \frac{\Lt_k(\kappa_1-2\kappa_2\Lt_k)}{\kappa_4(1-3\lambda_k)m^4} }
\end{equation}
and can be used to evaluate the partial fraction decomposition of $(1+w_0)^{-1}$. 
We then follow the lines of the transverse traceless part and resum the Matsubara modes. The end of Appendix \ref{app:scalarTrace}
shows that terms of the following form appear on the right hand side of the flow equation
\begin{align}
\Psi_{l,s}^{p,r} &= \sum_{n = - \infty}^\infty \frac{(m^2 n^2)^r}{\left( \Lt_k - (1 - \lambda_k)m^2n^2 \right)^s} \, \Phi_{l}^{p,q}(w_0) \, , \nonumber \\
\tilde{\Psi}_{l,s}^{p,r} &= \sum_{n = - \infty}^\infty \frac{(m^2 n^2)^r}{\left( \Lt_k - (1 - \lambda_k)m^2n^2 \right)^s} \, \tilde{\Phi}_{l}^{p,q}(w_0) \; .
\end{align}
Starting with the easiest one, we can 
sum the threshold function $\Phi^{1,0}_l(w_0)$. The result, with all auxiliary parameters set to one, reads
\begin{equation}
\Psi^{1,0}_{l,0} = \tfrac{1}{\Gamma(1+l)}\tfrac{1}{1-3\lambda_k}\tfrac{\pi}{m^4(w_+-w_-)}\left[ \tfrac{\Lt_k - (1 - \lambda_k)m^2w_-}{\sqrt{w_-}\tan(\pi\sqrt{w_-})} - \tfrac{\Lt_k - (1 - \lambda_k)m^2w_+}{\sqrt{w_+}\tan(\pi\sqrt{w_+})} \right] \; .
\end{equation}
However, we need further summations, which are more complicated. Investigating the result at the end of Appendix \ref{app:scalarTrace} we find that the following combinations are needed: $(p,r,s)=(1,0,1)$, $(2,0,1)$, $(2,1,1)$, $(2,1,2)$, $(2,2,2)$, $(3,1,2)$, $(3,2,2)$, $(3,3,2)$. To find all these terms we start with $\Psi^{1,0}_{l,1}$. This can be found by taking one derivative with respect to the parameter $\kappa_5$. Afterwards we set it to one, since we will not need it again. We use $\frac{1}{\Lt_k}\partial_{\kappa_5}\Psi^{1,0}_{l,0} = \Psi^{1,0}_{l,1}$ to find
\begin{equation}
\Psi^{1,0}_{l,1} = \tfrac{1}{\Gamma(1+l)}\tfrac{1}{1-3\lambda_k}\tfrac{\pi}{m^4(w_+-w_-)}\left[ \tfrac{1}{\sqrt{w_-}\tan(\pi\sqrt{w_-})} - \tfrac{1}{\sqrt{w_+}\tan(\pi\sqrt{w_+})} \right] \; .
\end{equation}
For all further terms we can use the following rules, which use the other auxiliary parameters $\kappa_1,\ldots,\kappa_4$
\begin{align}\label{recrel2}
\Psi^{p+1,r}_{l,s} &= -\frac{1}{p}\partial_{\kappa_1}\Psi^{p,r}_{l,s} \, ,& \qquad  \Psi^{p+1,r+1}_{l,s+1} &= \frac{1}{p}\frac{1}{(1-6\lambda_k)\Lt_k}\partial_{\kappa_3}\Psi^{p,r}_{l,s} \, , \nonumber \\
\Psi^{p+1,r}_{l,s+1} &= \frac{1}{2p\Lt_k^2}\partial_{\kappa_2}\Psi^{p,r}_{l,s} \, ,& \qquad \Psi^{p+1,r+2}_{l,s+1} &= -\frac{1}{p}\frac{1}{1-3\lambda_k}\partial_{\kappa_4}\Psi^{p,r}_{l,s} \; .
\end{align}
The explicit form of the threshold functions are given in Appendix \ref{app:C}.
Similar to the transverse traceless sector we note that $\tilde\Psi_{l,s}^{p,r} = \frac{\Gamma(l+1)}{\Gamma(l+2)}\Psi_{l,s}^{p,r}$. Furthermore, for convenience, we introduce the summed version of the $q$ functions which we denote by
\begin{equation} \label{eq:summedqScalar}
S^{p,r}_{l,s} = \Psi^{p,r}_{l,s} - \tfrac{1}{2} \, \eta_N \, \tilde{\Psi}^{p,r}_{l,s} \, .
\end{equation}

In Section \ref{sec:betas} the flow equations are given in terms of the summed $q$ functions \eqref{eq:summedq2T} and \eqref{eq:summedqScalar}. This is a very compact notation and makes the beta functions readable.

\section{Explicit form of the threshold functions}\label{app:C}
For completeness and further referencing the explicit
form of the threshold functions obtained from the recursion relations
\eqref{recrel2} are collected in this appendix.

\be
\begin{split}
\Psi^{2,0}_{l,1}   = & \, -\tfrac{1}{\Gamma(1+l)} \tfrac{\pi}{2m^8(1-3\lambda_k)^2 (w_+ - w_-)^3} \times \\
& \, \bigg[ \tfrac{m^2w_-(\lambda_k-1)(3w_-+w_+) + \Lt_k(5w_--w_+)}{w_-^{3/2}\tan(\pi\sqrt{w_-})} - \tfrac{m^2w_+(\lambda_k-1)(3w_++w_-) + \Lt_k(5w_+-w_-)}{w_+^{3/2}\tan(\pi\sqrt{w_+})} \\
& \quad + \tfrac{\pi(w_--w_+)(m^2w_-(\lambda_k-1) + \Lt_k)}{w_-\sin(\pi\sqrt{w_-})^2} + \tfrac{\pi(w_--w_+)(m^2w_+(\lambda_k-1) + \Lt_k)}{w_+\sin(\pi\sqrt{w_+})^2} \bigg] \, ,
\end{split}
\ee
\be
\begin{split}
\Psi^{2,1}_{l,1} = & \, \tfrac{1}{\Gamma(1+l)} \tfrac{\pi}{2m^6(1-3\lambda_k)^2 (w_- - w_+)^3} \times \\
& \, \bigg[ \tfrac{m^2w_-(3w_++w_-)(\lambda_k-1) + (3w_-+w_+)\Lt_k}{\sqrt{w_-}\tan(\pi\sqrt{w_-})} - \tfrac{m^2w_+(w_++3w_-)(\lambda_k-1) + (3w_++w_-)\Lt_k}{\sqrt{w_+}\tan(\pi\sqrt{w_+})}  \\
& \quad  + \tfrac{\pi(w_--w_+)(m^2w_-(\lambda_k-1) + \Lt_k)}{\sin(\pi\sqrt{w_-})^2} + \tfrac{\pi(w_--w_+)(m^2w_+(\lambda_k-1) + \Lt_k)}{\sin(\pi\sqrt{w_+})^2} \bigg] \, ,
\end{split}
\ee
\be
\begin{split}
\Psi^{2,1}_{l,2} = & \, \tfrac{1}{\Gamma(1+l)} \tfrac{\pi}{2m^6(1-3\lambda_k)^2(w_--w_+)^3} \times \\
& \, \bigg[ \tfrac{3w_-+w_+}{\sqrt{w_-}\tan(\pi\sqrt{w_-})} - \tfrac{w_-+3w_+}{\sqrt{w_+}\tan(\pi\sqrt{w_+})} 
- \pi \left(w_+ - w_- \right) \left( \tfrac{1}{\sin(\pi\sqrt{w_+})^2} + \tfrac{1}{\sin(\pi\sqrt{w_-})^2} \right)\bigg] \, , 
\end{split}
\ee
\be
\begin{split}
\Psi^{2,2}_{l,2} = & \, \tfrac{1}{\Gamma(1+l)} \tfrac{\pi}{2m^4(1-3\lambda_k)^2 (w_- - w_+)^3} \times \\
& \, \bigg[ \tfrac{\sqrt{w_-}(w_-+3w_+)}{\tan(\pi\sqrt{w_-})} - \tfrac{\sqrt{w_+}(3w_-+w_+)}{\tan(\pi\sqrt{w_+})} + \pi(w_--w_+) \left( \tfrac{w_-}{\sin(\pi\sqrt{w_-})^2} + \tfrac{w_+}{\sin(\pi\sqrt{w_+})^2} \right) \bigg] \, ,
\end{split}
\ee
\be
\begin{split}
\Psi^{3,1}_{l,2} = & \, -\tfrac{1}{\Gamma(1+l)} \tfrac{\pi}{8m^{10}(1-3\lambda_k)^3(w_--w_+)^5} \times \\
& \, \bigg[ -\tfrac{3m^2w_+(\lambda_k-1)(w_-^2+10w_-w_++5w_+^2) - \Lt_k(w_-^2-14w_-w_+-35w_+^2)}{w_+^{3/2}\tan(\pi\sqrt{w_+})}  \\
& \quad +\tfrac{3m^2w_-(\lambda_k-1)(5w_-^2+10w_-w_++w_+^2) + \Lt_k(35w_-^2+14w_-w_+-w_+^2)}{w_-^{3/2}\tan(\pi\sqrt{w_-})}  \\
& \quad - \tfrac{2\pi^2(w_--w_+)^2(m^2w_+(\lambda_k-1) + \Lt_k)}{\sqrt{w_+}\tan(\pi\sqrt{w_+})\sin(\pi\sqrt{w_+})^2}+ \tfrac{2\pi^2(w_--w_+)^2(m^2w_-(\lambda_k-1) + \Lt_k)}{\sqrt{w_-}\tan(\pi\sqrt{w_-})\sin(\pi\sqrt{w_-})^2}  \\
& \quad + \tfrac{\pi(w_--w_+)(m^2w_+(5w_-+7w_+)(\lambda_k-1) + \Lt_k(w_-+11w_+))}{w_+\sin(\pi\sqrt{w_+})^2} \\ 
& \quad + \tfrac{\pi(w_--w_+)(m^2w_-(7w_-+5w_+)(\lambda_k-1) + \Lt_k(11w_-+w_+))}{w_-\sin(\pi\sqrt{w_-})^2} \bigg] \, ,
\end{split}
\ee
\be
\begin{split}
\Psi^{3,2}_{l,2} = & \, \tfrac{1}{\Gamma(1+l)} \tfrac{\pi}{8m^8(1-3\lambda_k)^3(w_--w_+)^5} \times \\
& \, \bigg[ (m^2w_-(\lambda_k-1)+\Lt_k)\Big( \tfrac{12\sqrt{w_+}(w_-+w_+)}{\tan(\pi\sqrt{w_+})} - \tfrac{3(w_-^2+6w_-w_++w_+^2)}{\sqrt{w_-}\tan(\pi\sqrt{w_-})}   \\
& \qquad \qquad - \tfrac{2\pi^2\sqrt{w_-}(w_--w_+)^2}{\tan(\pi\sqrt{w_-})\sin(\pi\sqrt{w_-})^2}  - \pi(w_--w_+)\Big( \tfrac{3w_-+5w_+}{\sin(\pi\sqrt{w_-})^2} + \tfrac{4w_+}{\sin(\pi\sqrt{w_+})^2} \Big)\Big)  \\
& \;  -(m^2w_+(\lambda_k-1)+\Lt_k) \Big( \tfrac{12\sqrt{w_-}(w_-+w_+)}{\tan(\pi\sqrt{w_-})} - \tfrac{3(w_-^2+6w_-w_++w_+^2)}{\sqrt{w_+}\tan(\pi\sqrt{w_+})}  \\
& \qquad \qquad  - \tfrac{2\pi^2\sqrt{w_+}(w_--w_+)^2}{\tan(\pi\sqrt{w_+})\sin(\pi\sqrt{w_+})^2} +\pi(w_--w_+)\Big( \tfrac{5w_-+3w_+}{\sin(\pi\sqrt{w_+})^2} + \tfrac{4w_-}{\sin(\pi\sqrt{w_-})^2} \Big)\Big)\bigg] \, ,
\end{split}
\ee
\be
\begin{split}
\Psi^{3,3}_{l,2} = & \, -\tfrac{1}{\Gamma(1+l)} \tfrac{\pi}{8m^6(1-3\lambda_k)^3(w_--w_+)^5} \times \\
& \, \bigg[ \tfrac{\sqrt{w_-}(-m^2w_-(w_-^2-14w_-w_+-35w_+^2)(\lambda_k-1) + 3\Lt_k(w_-^2+10w_-w_++5w_+^2))}{\tan(\pi\sqrt{w_-})}  \\
& \, + \tfrac{\sqrt{w_+}(m^2w_+(w_+^2-14w_-w_+-35w_-^2)(\lambda_k-1) - 3\Lt_k(w_+^2+10w_-w_++5w_-^2))}{\tan(\pi\sqrt{w_+})}  \\
& \, + \tfrac{2\pi^2w_-^{3/2}(w_--w_+)^2(m^2w_-(\lambda_k-1) + \Lt_k)}{\tan(\pi\sqrt{w_-})\sin(\pi\sqrt{w_-})^2} - \tfrac{2\pi^2w_+^{3/2}(w_--w_+)^2(m^2w_+(\lambda_k-1) + \Lt_k)}{\tan(\pi\sqrt{w_+})\sin(\pi\sqrt{w_+})^2} \\
& \, + \pi(w_--w_+)\Big( \tfrac{w_-(m^2w_-(13w_+-w_-)(\lambda_k-1) + 3\Lt_k(w_-+3w_+))}{\sin(\pi\sqrt{w_-})^2} \\
& \qquad \qquad \qquad \qquad + \tfrac{w_+(m^2w_+(13w_--w_+)(\lambda_k-1) + 3\Lt_k(3w_-+w_+))}{\sin(\pi\sqrt{w_+})^2} \Big) \bigg] \, .
\end{split}
\ee

\end{appendix}

\end{document}